\pdfoutput=1

\documentclass[11pt,twoside,a4paper,cmspaper,final,collab]{cms-tdr}
\def\svnVersion{444261P}\def\cmsCernNoTag{CERN-EP-2017-144}\def\cmsCernDate{\today}\def\cmsMessage{Published in Physics Letters B as \href{http://dx.doi.org/10.1016/j.physletb.2018.01.012}{\doi{10.1016/j.physletb.2018.01.012.}}}

\begin{document}\cmsNoteHeader{SUS-16-032}

\hyphenation{had-ron-i-za-tion}
\hyphenation{cal-or-i-me-ter}
\hyphenation{de-vices}
\RCS$HeadURL: svn+ssh://svn.cern.ch/reps/tdr2/papers/SUS-16-032/trunk/SUS-16-032.tex $
\RCS$Id: SUS-16-032.tex 441916 2018-01-18 17:43:55Z safarzad $

\newlength\cmsFigWidth
\ifthenelse{\boolean{cms@external}}{\setlength\cmsFigWidth{0.49\textwidth}}{\setlength\cmsFigWidth{0.45\textwidth}}
\ifthenelse{\boolean{cms@external}}{\providecommand{\cmsLeft}{upper\xspace}}{\providecommand{\cmsLeft}{left\xspace}}
\ifthenelse{\boolean{cms@external}}{\providecommand{\cmsRight}{lower\xspace}}{\providecommand{\cmsRight}{right\xspace}}

\newcommand{\TQCD}{\ensuremath{T_{\mathrm{QCD}}}\xspace}
\newcommand{\mct}{\ensuremath{M_{\mathrm{CT}}}\xspace}
\newcommand{\dphimet}{\ensuremath{\Delta\phi_{\mathrm{min}}}\xspace}
\newcommand{\HTb}{\ensuremath{H_{\mathrm{T}}^{\mathrm{b}}}\xspace}
\newcommand{\HTc}{\ensuremath{H_{\mathrm{T}}^{\mathrm{c}}}\xspace}
\newcommand{\Nb}{\ensuremath{N_{\text{b--tags}}}\xspace}
\newcommand{\NI}{\ensuremath{N_{\text{SV}}}\xspace}
\newcommand{\Nc}{\ensuremath{N_{\text{c--tags}}}\xspace}
\newcommand{\minMT}{\ensuremath{M_{\mathrm{T}}^{\mathrm{min}}(\pt(\mathrm{j}_{1,2}),\met)}\xspace}
\newcommand{\ptvmiss}{\ensuremath{\vec{p}_{\mathrm{T}}^{\text{miss}}}\xspace}
\newcommand{\ptv}{\ensuremath{\vec{p}_{\mathrm{T}}}\xspace}
\newcommand{\met}{\ensuremath{p_{\mathrm{T}}^{\text{miss}}}\xspace}
\newcommand{\metim}{\ensuremath{|(\vec{p}_{\mathrm{T}}(\text{ISR})+\ptvmiss)|/\met}\xspace}
\newcommand{\zll}{\ensuremath{\PZ\to \ell^+\ell^-}\xspace}
\newcommand{\zjets}{\ensuremath{\PZ+}jets\xspace}
\newcommand{\wjets}{\ensuremath{\PW+}jets\xspace}
\newcommand{\znunu}{\ensuremath{\PZ\to\PGn\PAGn}\xspace}
\newcommand{\fulllumi}{35.9\fbinv}
\newcommand{\stopq}{\ensuremath{\PSQt_{1}}\xspace}
\newcommand{\stopqbar}{\ensuremath{\PASQt_{1}}\xspace}
\newcommand{\sbottomq}{\ensuremath{\PSQb_{1}}\xspace}
\newcommand{\sbottomqbar}{\ensuremath{\PASQb_{1}}\xspace}
\newcommand{\lsp}{\PSGczDo}
\newcommand{\ttZ}{\ensuremath{\ttbar\Z}\xspace}
\newcommand{\ttW}{\ensuremath{\ttbar\PW}\xspace}
\newcommand{\WZ}{\ensuremath{\PW\PZ}\xspace}
\newcommand{\WW}{\ensuremath{\PW\PW}\xspace}
\newcommand{\ZZ}{\ensuremath{\PZ\PZ}\xspace}
\newcommand{\x}{\ensuremath{\phantom{0}}}
\cmsNoteHeader{SUS-16-032}

\title{Search for the pair production of third-generation squarks with two-body decays to a bottom or charm quark and a neutralino in proton-proton collisions at \texorpdfstring{$\sqrt{s}=13\TeV$}{sqrt(s) = 13 TeV}}

\address[ipm]{School of Particles and Accelerator Inst. for Res. in Fundam. Sc.(IPM)
}

\date{\today}

\abstract{
Results are presented from a search for the pair production of third-generation squarks in proton-proton collision events with two-body decays to bottom or charm quarks and a neutralino, which produces a significant imbalance in the transverse momentum. The search is performed using a sample of proton-proton collision data at $\sqrt{s}=13\TeV$ recorded by the CMS experiment at the LHC, corresponding to an integrated luminosity of 35.9\fbinv. No statistically significant excess of events is observed beyond the expected contribution from standard model processes. Exclusion limits are set in the context of simplified models of bottom or top squark pair production. Models with bottom squark masses up to 1220\GeV are excluded at 95\% confidence level for light neutralinos, and models with top squark masses of 510\GeV are excluded assuming that the mass splitting between the top squark and the neutralino is small. 	
}

\hypersetup{%
pdfauthor={CMS Collaboration},%
pdftitle={Search for direct production of bottom and top squark pairs in proton-proton collisions at sqrt(s) = 13 TeV},%
pdfsubject={CMS},%
pdfkeywords={CMS, physics, supersymmetry}}

\maketitle

\section{Introduction}
\label{sec:intro}

The standard model (SM) has been extremely successful in describing
particle physics phenomena. Nevertheless, it suffers from shortcomings such as the hierarchy problem~\cite{Barbi}, the need for a fine-tuned cancellation of large quantum corrections to the Higgs mass to maintain a physical value at the observed electroweak scale.
Supersymmetry
(SUSY)~\cite{Ramond,Golfand,Neveu,Volkov,Wess,Wess1,Fayet,Nilles} postulates a symmetry between bosons and fermions and provides a ``natural'' solution to the hierarchy problem through the cancellation of quadratic divergences in particle and SUSY particle loop corrections to the Higgs boson mass.
In natural SUSY models, light top and bottom squarks are preferred with masses close to the electroweak scale~\cite{Barbi,Pap}.
In $R$-parity conserving SUSY models~\cite{Farrar:1978xj}, SUSY particles are created in pairs, and the lightest SUSY particle (LSP) is stable. The LSP is assumed here to be the lightest neutralino (\lsp), which is both weakly interacting and stable and therefore has the properties of a dark matter candidate~\cite{darkmatter}.

This letter presents searches for the direct production of pairs of bottom
($\sbottomq\sbottomqbar$) and top ($\stopq\stopqbar$) squarks, decaying
to multijet final states with a large transverse momentum imbalance. The search is performed using~\fulllumi of data collected in proton-proton ($\Pp\Pp$)
collisions by the CMS detector, at a centre-of-mass energy of 13\TeV, at the CERN LHC~\cite{LHC_jinst}.

\begin{figure}[hbtp]
\centering
\includegraphics[width=0.40\textwidth]{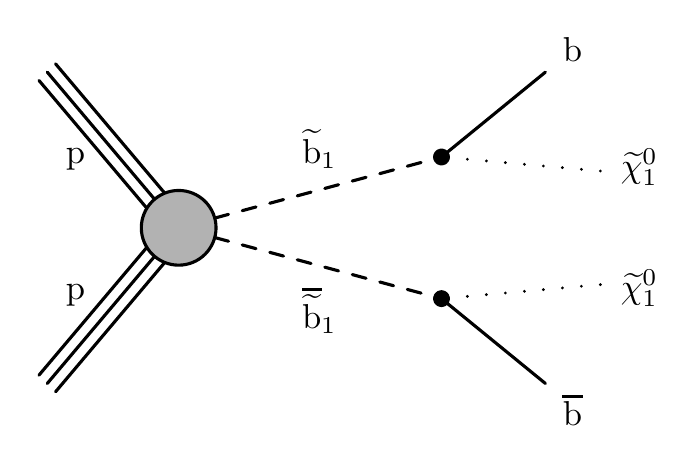}
\includegraphics[width=0.40\textwidth]{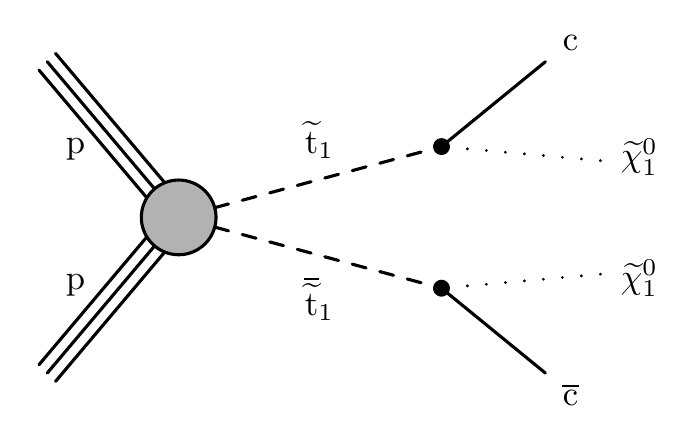} \\
\caption{Diagrams showing the pair production of bottom or top squarks followed by their decays according to $\sBot\to \cPqb \chiz_1$ (\cmsLeft)
and $\sTop\to \PQc\chiz_1$ (\cmsRight).
}
\label{fig:T2bb}
\end{figure}

The search for bottom squark pair production is based on the decay mode $\sbottomq\to\PQb\lsp$.
This study considers a scenario for top-squark decay that can arise when the mass splitting, $\Delta m \equiv m_{\stopq}-m_{\lsp} $ is below the mass of the \PW~boson. The decay process $\stopq\to\PQt\lsp, \PQt\to\PQb\PW$ is then suppressed not only because the top quark must be virtual, but also because the W boson must be virtual as well. If flavor-changing neutral current decays $\stopq\to\PQc\lsp$ are allowed, then the branching fraction for the two-body decay $\stopq\to\PQc\lsp$ can in principle become substantial.
Bottom and top squark pair productions are studied in the context of simplified models \cite{Simp1,Simp2,Simp3}.
Figure~\ref{fig:T2bb} illustrates the bottom and top squark decay modes explored in this  letter.

The search techniques are based on the work presented in Ref.~\cite{SUS16008} but use
improved discrimination tools to exploit specific kinematic characteristics
of the signal models. A charm quark tagging algorithm is used
in the top squark search to identify c quarks originating from top
squark decays. In addition, specific object reconstruction tools are
employed to improve sensitivity to compressed spectrum scenarios, where visible decay products carry low momenta.
The new methods and discriminators, as well as the increase in integrated luminosity, lead to considerably improved sensitivity relative to previous searches.While the analysis improvement for compressed spectra is due to the charm and soft b quark identification, the increase in the luminosity provides the improved sensitivity for the noncompressed spectra. Results
of similar searches were previously reported by the ATLAS and CMS
Collaborations, using pp collisions at 7, 8, and 13
\TeV~\cite{ATLAS1,ATLAS2,ATLAS5,ATLAS5a,ATLAS6,ATLAS7,ATLAS8,atlas-stop1l-2015,CMS-STOP-lepton,CMS-alphaT,RAZOR_8TeV,stop8TeV,stop0l_8TeV,SUS15002,SUS15003,SUS15004,SUS15005,SUS16033,ATLASCom,ATLAS2b,ATLASJet}, as well as by the CDF and D0 Collaborations in proton-antiproton collisions at the Fermilab Tevatron~\cite{CDF1,CDF2,D01,D02}.

\section{The CMS detector}
\label{sec:detector}

The central feature of the CMS detector is a superconducting solenoid of 6\unit{m} internal diameter,
providing a magnetic field of 3.8\unit{T}. An all-silicon pixel and strip tracker, a
lead tungstate crystal electromagnetic calorimeter (ECAL), and a brass and scintillator hadron calorimeter, each composed of
a barrel and two endcap sections are located within the field volume.
Forward calorimeters extend the pseudorapidity ($\eta$) coverage provided by the barrel and endcap detectors.
Muons are measured in gas-ionization detectors embedded in the steel flux-return yoke outside the solenoid.
The first level of the CMS trigger system, composed of specialized hardware processors, uses information from the
calorimeters and muon detectors to select the most interesting events in a fixed time interval of less than 4\mus. A high level
trigger processor farm decreases the event rate from around 100\unit{kHz} to less than 1\unit{kHz}, before data storage~\cite{Trigger}.
A more detailed description of the CMS detector, together with a definition of the coordinate system and relevant
kinematic variables, can be found in Ref.~\cite{CMS_jinst}.

\section{Event reconstruction and Monte Carlo simulation}
\label{sec:reco}

Events are reconstructed with the particle flow (PF) algorithm~\cite{PF}, which combines information from the
subdetectors to optimize reconstruction and identification of produced stable particles, namely charged and neutral hadrons, photons, electrons, and
muons. Events selected for this search are required to pass filters designed to remove detector- and beam-related noise and must have at least one reconstructed vertex. Usually more than one such vertex is reconstructed, due to pileup, \ie multiple pp collisions within
the same or neighbouring bunch crossings. The reconstructed vertex with the largest value of summed physics-object $\pt^2$ is taken to be the primary $\Pp\Pp$ interaction vertex (PV), where \pt is the transverse momentum. The physics objects are the objects returned by a jet finding algorithm~\cite{Cacciari:2008gp,Cacciari:2011ma} applied to all charged tracks associated with the vertex, plus the corresponding associated missing transverse momentum.

Charged particles originating from the primary vertex, photons, and neutral hadrons are clustered into jets using the anti-\kt algorithm~\cite{Cacciari:2008gp} implemented in \FASTJET~\cite{Cacciari:2011ma} with a distance parameter of 0.4. The jet energy is corrected for the contribution from pileup based on the jet area method~\cite{Jet2}. Additional corrections to the jet energy scale are applied to compensate for variations in detector response~\cite{JER}. Jets are required to have \pt greater than 25\GeV and to be contained within the tracker volume, $\abs{\eta} < 2.4$.
The momentum imbalance vector (\ptvecmiss) is calculated
as the negative vector sum of transverse momenta of all PF candidates reconstructed in an event, and its magnitude is referred to as missing transverse momentum, denoted \met~\cite{Met1}.

Muons are reconstructed by combining the information from the silicon tracker and the muon detectors in a global fit.
An identification selection is performed using the quality of the geometrical matching between the tracker and the muon system measurements~\cite{Muon}.
Electron candidates are reconstructed by matching clusters of energy deposited in the
ECAL to reconstructed tracks. Selection criteria based on the distribution of the shower shape,
track cluster matching, and consistency between the cluster energy and track momentum are
then used in the identification of electron candidates~\cite{Khachatryan:2015hwa}. Muon and electron
candidates are required to have $\pt > 10\GeV$, to be within $|\eta|< 2.4$, and to originate from within 2~mm of the beam axis in the transverse plane.
Relative lepton isolation, $I_\text{rel}$,  is quantified as the sum of the \pt of PF candidates within a cone $\Delta R =\sqrt{\smash[b]{(\Delta\eta)^2+(\Delta\phi)^2}}$ around the lepton (where $\phi$ is the azimuthal angle in radians), divided by the lepton \pt. The lepton itself and charged PF candidates not originating from the PV are not considered in the sum. The isolation sum is corrected for effects of pileup interactions through an area-based estimate~\cite{CMS-PAS-JME-14-001} of the pileup energy deposited in the cone. The size of the cone is defined according to lepton \pt, as follows:
\begin{equation}
\label{eqn:miniiso}
\DR =
\begin{cases}
0.2,          & \text{if } \pt < 50\GeV, \\
10\GeV/\pt,   & \text{if } 50 < \pt < 200\GeV, \\
0.05,         & \text{if } \pt > 200\GeV.
\end{cases}
\end{equation}
The shrinking cone radius for higher-\pt leptons maintains high efficiency for the
collimated decay products of highly-boosted heavy objects.

Jets are identified as b tagged using the combined secondary vertex (CSVv2) algorithm~\cite{Btags,BTV1}.
The b quark jet (``b jet") identification efficiencies
for jets with $\pt > 25\GeV$ and $\abs{\eta}< 2.4$ vary with jet \pt and are 80--85\% and 46--74\% for the loose and medium working points used in this analysis, respectively.  The probability for light-flavour (charm) jets to be mistagged as function of jet \pt is 8--12\% (40\%) for the loose working point and 1--2\% (20\%) for medium working point. The single muon \ttbar events are used to extract the charm mistag rate of the CSVv2 algorithm~\cite{BTV1}.

 A c quark tagging algorithm is used to identify jets originating from charm quarks (``c jets"), while rejecting either b or light-flavour jets~\cite{ctag}. Two classifiers are introduced, one to discriminate c jets from light-flavour, and one for discriminating
c jets from b jets. To identify c jets, a selection is implemented in the plane of the two discriminators. As c-jet properties are often distributed in between those of b- and light-jets, the charm tagger discriminators are less efficient than b-tagger and usually suffers from large misidentification rates. We get the best analysis sensitivity using the ``medium" working point version of the algorithm, which has 40\% c quark identification efficiency
for jets with $ \pt > 25\GeV$ and $|\eta|< 2.4$. The rate for b and light-flavour jets to be mistagged as a c jet is 20\%. The efficiency to identify c jets is measured with a sample enriched in c jets using events with a W boson produced in association with a c quark.

For the very compressed spectra (${m}_{\sbottomq}-m_{\lsp} < 25\GeV$), a large fraction of events contain \PQb~quarks with $ \pt < 25\GeV$, which may fail to pass the jet selection or the b tagging working points. We therefore extend the identification of \PQb~quarks based on the presence of a secondary vertex (SV) reconstructed using the inclusive vertex finder (IVF) algorithm~\cite{Khachatryan:2011wq}. To suppress the background originating from light-flavour jets, the following requirements are placed on the SV observables:
the distance in the transverse plane between the SV and PV must be ${<}3\unit{cm}$; there must be ${>}2$ tracks associated with the SV; the significance of this distance is required to be ${>} 4$; the cosine of the pointing angle, which is defined through the scalar product between the distance vector $(\overrightarrow{{\mathrm{SV},\mathrm{PV}}})$ and the $\vec{p}_{\mathrm{SV}}$ direction has to be ${>} 0.98$, where $\vec{p}_{\mathrm{SV}}$ is the total three-momentum of the tracks
associated with the SV. Finally, in order to avoid overlaps with the b and c tagging selections described above, the
distance $\Delta R$ of the SV to jets (including $\PQb$- or $\PQc$-tagged jets) has to be ${>} 0.4$, and the transverse component of $p_{\mathrm{SV}}$ is required to satisfy
$p_{\mathrm{SV}} < 25\GeV$. The method has 20\% efficiency in identifying  b hadrons versus less than one percent of misidentification and the performance in simulation agrees with the performance with data within 16\%~\cite{CMS-SUS-16-049}.

The Monte Carlo (MC) simulation of events is used to study the properties of SM backgrounds and signal models. The \MGvATNLO 2.2.2 generator~\cite{Alwall:2014hca} is used in leading-order (LO) mode to simulate events originating from \ttbar, \wjets, \zjets, and quantum chromodynamics multijet processes~('QCD'), as well as signal events, based on LO NNPDF3.0~\cite{Ball:2014uwa} parton distribution functions (PDFs). The LO MC is used for these SM processes because it allows a better control of the associated jet production to large multiplicities, while any next-to-leading order (NLO) MC would only model the first radiation at NLO and then use parton shower for extra jets. Single top quark events produced in the $\PQt\PW$ channel are generated at NLO with \textsc{Powheg} v2~\cite{Nason:2004rx,Frixione:2007vw,Alioli:2010xd,Re:2010bp}, while SM processes such as \WZ, \ZZ, \WW, \ttZ, and \ttW, which are grouped together as the rare processes because of the small contribution in this analysis, are generated at NLO using the~\MADGRAPH{}5\_a\MCATNLO 2.2.2 program, using NLO NNPDF3.0 PDFs. Parton showering and hadronization is generated using \textsc{Pythia}8.212~\cite{Sjostrand:2014zea}. The response of the CMS detector for the SM backgrounds is simulated with the \GEANTfour~\cite{geant4} package. The CMS fast simulation package~\cite{fastsim} is used to simulate all signal samples, and is verified to provide results that are consistent with those obtained from the full \GEANTfour-based simulation. Any residual differences in the detector response description between the \GEANTfour and fast simulation are corrected for, with corresponding uncertainties in the signal acceptance taken into account.
Event reconstruction is performed in the same manner as for collision data. A distribution of pileup interactions is used when producing the simulated samples. The samples are then reweighted to match the pileup profile observed in the collected data. The signal production cross sections are calculated using NLO with next-to-leading logarithm (NLL) soft-gluon resummation calculations~\cite{Borschensky:2014cia}. The most precise cross section calculations are used to normalize the SM simulated samples, corresponding most often to next-to-next-to-leading order (NNLO) accuracy.

\section{Event selection}
\label{sec:sel}

The recorded events are required to have
$\met > 100\GeV$
at the trigger level. To ensure full trigger efficiency, events selected offline are required to have $\met > 250\GeV$, as well as two, three, or four jets.
For bottom squark production, only two jets are expected from squark decays. For the
model involving top squarks with a small mass difference relative to the LSP, most decay products have small \pt and therefore the analysis relies on the presence of one or two additional jets from initial-state radiation (ISR). In both cases, the number of high-\pt jets is expected to be small, and therefore events with a
fifth jet with \pt above 75\GeV are rejected.  The event is discarded if it has more than five jets.

To reduce the SM background from processes with a leptonically decaying \PW~boson, we reject events containing isolated muons (electrons) with $I_\text{rel} <0.10$ ($I_\text{rel}< 0.21$). The contribution from hadronically decaying $\tau$ leptons ($\PGt_{\rm h}$) is reduced by placing a veto on events containing isolated charged-hadron PF candidates (isolated track) with $\pt > 10\GeV$, $|\eta| < 2.5$. Candidates are categorized
as being isolated if their isolation sum, \ie the scalar sum of the \pt of charged PF candidates
within a fixed cone of $R = 0.3$ around the candidate is smaller than 10\%
of the candidate \pt.

The dominant SM background sources are \zjets events with \znunu decay and background from \wjets, \ttbar, and single top quark processes with leptonic \PW~boson decays. These processes contribute
to the search regions when the lepton is not isolated or identified, or is out of kinematical or detector acceptance.
In addition, a hadronically decaying $\tau$ lepton can be reconstructed as a jet and hence
contributes to the signal region.
A smaller background contribution comes from QCD multijet events in which large \met originates from jet mismeasurements. The direction of \ptvecmiss in such events is often aligned with one of the mismeasured jets. To suppress this background, the absolute difference in the azimuthal angle ($\dphimet$) between \ptvecmiss and the closest of the three jets with highest (\ie leading)~\pt is required to be ${>} 0.4$.

Two sets of search regions are defined to optimize the sensitivity for signal with either compressed or noncompressed mass spectra.
In addition to the criteria discussed above, in models with noncompressed mass spectra we require the \pt of the leading jet to be ${>}100\GeV$
and to contain at least one additional jet with $\pt > 75\GeV$. We also require the two leading jets to be b tagged. These requirements suppress events originating from \PW~and \PZ~boson production, in which the leading jets have softer \pt spectra, as they are produced by ISR.
To maintain a stable \PQb~tagging efficiency as a function of jet \pt, both the loose and medium working points of the \PQb~tagging algorithm are used to identify \PQb~jets.  The \PQb~tagging efficiency of the medium working point depends strongly on the jet \pt and degrades by about 20--30\% for jets with $\pt > 500\GeV$, while the efficiency of the loose working point is more stable with increasing jet \pt. Specifically, we use the loose working point to identify a leading \PQb-tagged jet if it has $\pt > 500\GeV$, and otherwise use the medium working point.  Since such high-\pt jets are less likely to occur in SM processes, the higher misidentification rate of the loose working point provides only a small increase in the SM background. The third and fourth jet if present, are required to have  $\pt > 30\GeV$.

{\tolerance = 1200
In \ttbar events with a lost lepton, the transverse mass distribution of the neutrino and b quark from the same top quark decay has
an endpoint at the mass of the top quark. The observable~$\minMT$ is defined as
\begin{equation}
\ifthenelse{\boolean{cms@external}}{
\begin{split}
\minMT \equiv  &  \\ \text{min}[ M_{\mathrm{T}}(\pt(\text{j}_{1}),& \met), M_{\text{T}}(\pt(\text{j}_{2}), \met) ],
\end{split}
}
{
\minMT~\equiv~\text{min}[ M_{\mathrm{T}}(\pt(\text{j}_{1}), \met), M_{\text{T}}(\pt(\text{j}_{2}), \met) ],
}
\end{equation}
where
\ifthenelse{\boolean{cms@external}}{
\begin{equation*}
M_{\text{T}}(\pt(\text{j}_{1,2}), \met) = \sqrt{\smash[b]{2\pt(\text{j}_{1,2})(1-\cos\Delta\phi(\text{j}_{1,2},\met))}},
\end{equation*}
}
{
$M_{\text{T}}(\pt(\text{j}_{1,2}), \met) = \sqrt{\smash[b]{2\pt(\text{j}_{1,2})(1-\cos\Delta\phi(\text{j}_{1,2},\met))}}$,
}
$ \pt(\text{j}_{1}) $ and $\pt(\text{j}_{2})$ are the transverse momenta of the two leading jets, and $\Delta\phi(\text{j}_{1,2},\met)$ is the azimuthal angle between leading (sub-leading) jet and \ptvmiss. Imposing a minimum requirement of 250\GeV on \minMT reduces a significant portion of the \ttbar background.
\par}

Events in this sample are then categorized by \HT, defined as the scalar sum of the \pt of the two leading jets, and the boost-corrected contransverse mass~\cite{MCT1,MCT}, \mct, defined as:
\begin{equation}
\mct^2(j_1, j_2)
 =  2\pt(\text{j}_1)\pt(\text{j}_2) (1+\cos\Delta\phi(\text{j}_1,\text{j}_2)),
\end{equation}
where $\Delta\phi(\text{j}_1,\text{j}_2)$ is the azimuthal angle between two leading jets. For models in which particles are pair produced and have the same decay chain, the \mct distribution has an endpoint determined by the masses of the parent and daughter particles. For the decay $\sbottomq \to \PQb\lsp$, this endpoint is at mass  $(m_{\sbottomq}^{2}-m_{\tilde{{\chi}}_{1}^{0}}^2)/m_{\sbottomq}$. A  minimum requirement of 150\GeV on \mct is applied.

For signals with compressed mass spectra,
high-\pt ISR jet is required to reconstruct the decay chain of quarks as jets and to obtain a large value of \met.
Since such ISR jets are not expected to originate from \PQb or \PQc~quarks, the leading jet is required to fail the loose b tagging and medium c tagging requirements to define the ISR system according to whether the sub-leading jet is \PQb- or \PQc-tagged. If the next-to-leading jet \pt in the event is ${>}50\GeV$ and is neither \PQb- or \PQc-tagged, the ISR system is defined by the two leading jets; otherwise only the leading jet is considered as the ISR system. The ISR system \pt is required to exceed 250\GeV. The jet imbalance in the transverse plane is quantified as the vector sum of the ISR system \ptv and \ptvmiss, divided by \met, \metim. For the topology of
interest, the transverse momentum imbalance must be
small and we therefore require that $\metim < 0.5$.

The \PQb- or \PQc-tagged jet, using medium b and c tagging requirements,  must have $\pt > 25\GeV$, and if a \PQb-tagged jet is also identified as \PQc-tagged jet, it is only counted once as a \PQb-tagged jet.

The \mct observable loses its discriminating power in the compressed models when the mass splitting
between the parent particle and the $\lsp$ is small.
Therefore, we use as the main discriminants the number of \PQb- and \PQc-tagged jets (\Nb and \Nc, respectively) and a number of selected SVs (\NI) and \met. If there are at least one b- or c-tagged jets the extra variables, \HTb, and \HTc, which reflect the scalar sums of transverse momenta of \PQb- and \PQc-tagged jets, respectively, are used.
The search region with $\NI > 0 $ provides the sensitivity in the very compressed spectra for the bottom squark search.

The baseline selections in both the noncompressed and compressed regions are summarized in Table~\ref{tab:sel}, and
the signal region definitions in both regions are shown in Tables~\ref{tab:noncom_bin} and~\ref{tab:com_bin0}, respectively.

\begin{table*}[!ht]
\topcaption{A summary of the baseline selections used for the noncompressed and compressed search regions.}
\label{tab:sel}
\centering
\begin{tabular}{l  c  c}
 \multicolumn{1}{c  }{} & \multicolumn{2}{c}{Search regions \ \ \ \ \ \ \  }    \\
 \cline{2-3}
& Noncompressed  & Compressed   \\
 \hline
$N_{\text{jets}}$& 2--4 ($\pt> 30\GeV$) & 2--4 ($\pt> 25\GeV$)   \\
Jet veto&  $5^{\text{th}}\text{-jet}$ ($\pt > 75\GeV$) &  $5^{\text{th}}\text{-jet}$ ($\pt > 75\GeV$)\\
Lepton veto& e, $\mu$, and isolated track &  e, $\mu$, and isolated track    \\
Leading jet & \begin{tabular}{c} $\pt> 100\GeV$ \\ and is b tagged \end{tabular} & \begin{tabular}{c} $\pt> 100\GeV$\\ and is not \PQb or \PQc tagged \end{tabular}\\
Sub-leading jet & \begin{tabular}{c} $\pt> 75\GeV$ \\ and is b tagged \end{tabular}  & \begin{tabular}{c} $\pt> 25(50)\GeV$\\ and is (is not) \PQb or \PQc tagged \end{tabular} \\
\met&  ${>} 250\GeV$ &  ${>} 250\GeV$  \\
\pt(ISR) & --- &${>} 250\GeV$ \\
\dphimet & ${>} 0.4\unit{rad}$ & ${>} 0.4\unit{rad}$\\
${|(\vec{p}_{\mathrm{T}}(\text{ISR})+\ptvmiss)|}/{\met} $ & --- & ${<}0.5$ \\
\minMT &${>} 250\GeV$ & --- \\
\mct  &${>}150\GeV$& --- \\
\end{tabular}
\end{table*}

\begin{table*}[!ht]
\topcaption{The categorization of \HT and \mct for search regions in noncompressed signal models. }
\label{tab:noncom_bin}
\centering
\begin{tabular}{l| l}
\multicolumn{2}{c}{Noncompressed regions}    \\ \hline
\HT [\GeVns{}] &  \mct [\GeVns{}]    \\ \hline
200--500 & 150--250, 250--350, 350--450, $>$450 \\
500--1000 & 150--250, 250--350, 350--450, 450--600, $>$600 \\
$>$1000& 150--250, 250--350, 350--450, 450--600, 600--800, $>$800     \\
\end{tabular}
\end{table*}

\begin{table*}[!ht]
\caption{The categorization in \Nb, \Nc, \NI, \HT, and \met for search regions in models with compressed spectra. Only events with zero \PQb-tagged jets are used to define the search regions with exactly one or two c-tagged jets. }
\label{tab:com_bin0}
\centering
\begin{tabular}{c| c c}
\multicolumn{3}{c}{Compressed regions}    \\ \hline
\Nb, \Nc, \NI&    \met [\GeVns{}] & \HT (\PQb- or \PQc-tagged jets) [\GeVns{}]   \\ \hline
\multirow{5}{*}{\Nb = 1}& 250--300 &  $<$100  \\
& 300--500 &  $<$100  \\
& 500--750 &  $<$100  \\
& \x750--1000 &  $<$100  \\
& $>$1000 & $<$100 \\ \hline
\multirow{6}{*}{\Nb = 2}&  \multirow{2}{*}{250--300} & $<$100  \\
&  & 100--200  \\
& \multirow{2}{*}{300--500} & $<$100  \\
& & 100--200  \\
& \multirow{2}{*}{$>$500} & $<$100  \\
& & 100--200  \\ \hline
\multirow{5}{*}{\Nc = 1}& 250--300 &  $<$100  \\
& 300--500 &  $<$100  \\
& 500--750 &  $<$100  \\
& \x750--1000 &  $<$100  \\
& $>$1000 & $<$100  \\ \hline
\multirow{8}{*}{\Nc = 2}&  \multirow{2}{*}{250--300} & $<$100  \\
&  & 100--200  \\
& \multirow{2}{*}{300--500} & $<$100  \\
& & 100--200  \\
& \multirow{2}{*}{500--750} & $<$100  \\
& & 100--200  \\
& \multirow{2}{*}{$>$750} & $<$100 \\
& & 100--200  \\ \hline
\multirow{5}{*}{\Nb+ \Nc+ \NI = 0}& 300--500 & ---\\
& 500--750 & ---\\
& \x750--1000 & ---\\
& 1000--1250 & --- \\
& $>$1250 & ---\\ \hline
\multirow{5}{*}{\Nb+ \Nc = 0, $\NI > 0$}& 250--300 & ---\\
& 300--500 &  --- \\
& 500--750 & --- \\
& \x750--1000 & ---\\
& $>$1000 & ---\\
\end{tabular}
\end{table*}

\begin{figure}[!ht]
\centering
\includegraphics[width=\cmsFigWidth]{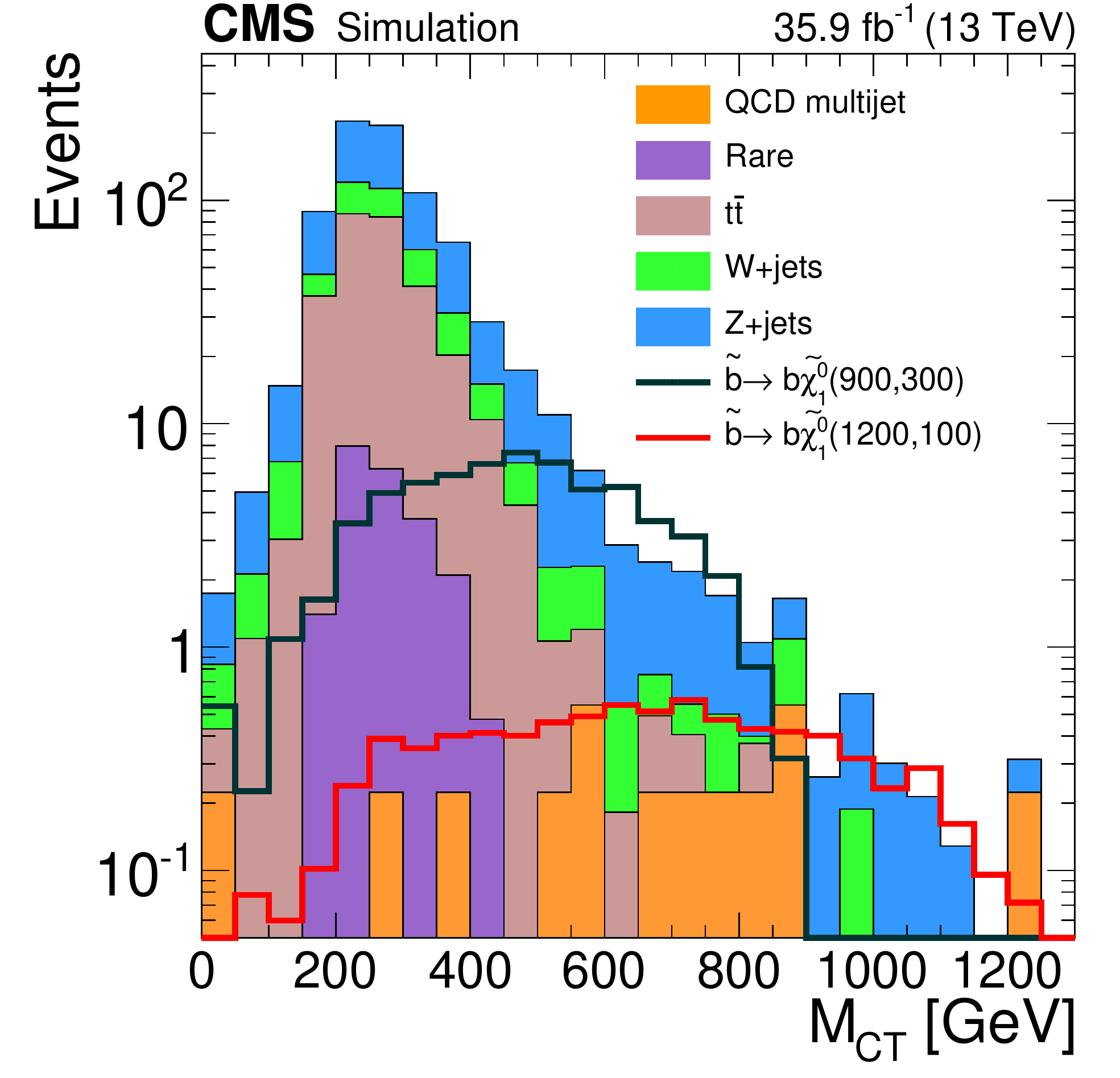}
\includegraphics[width=\cmsFigWidth]{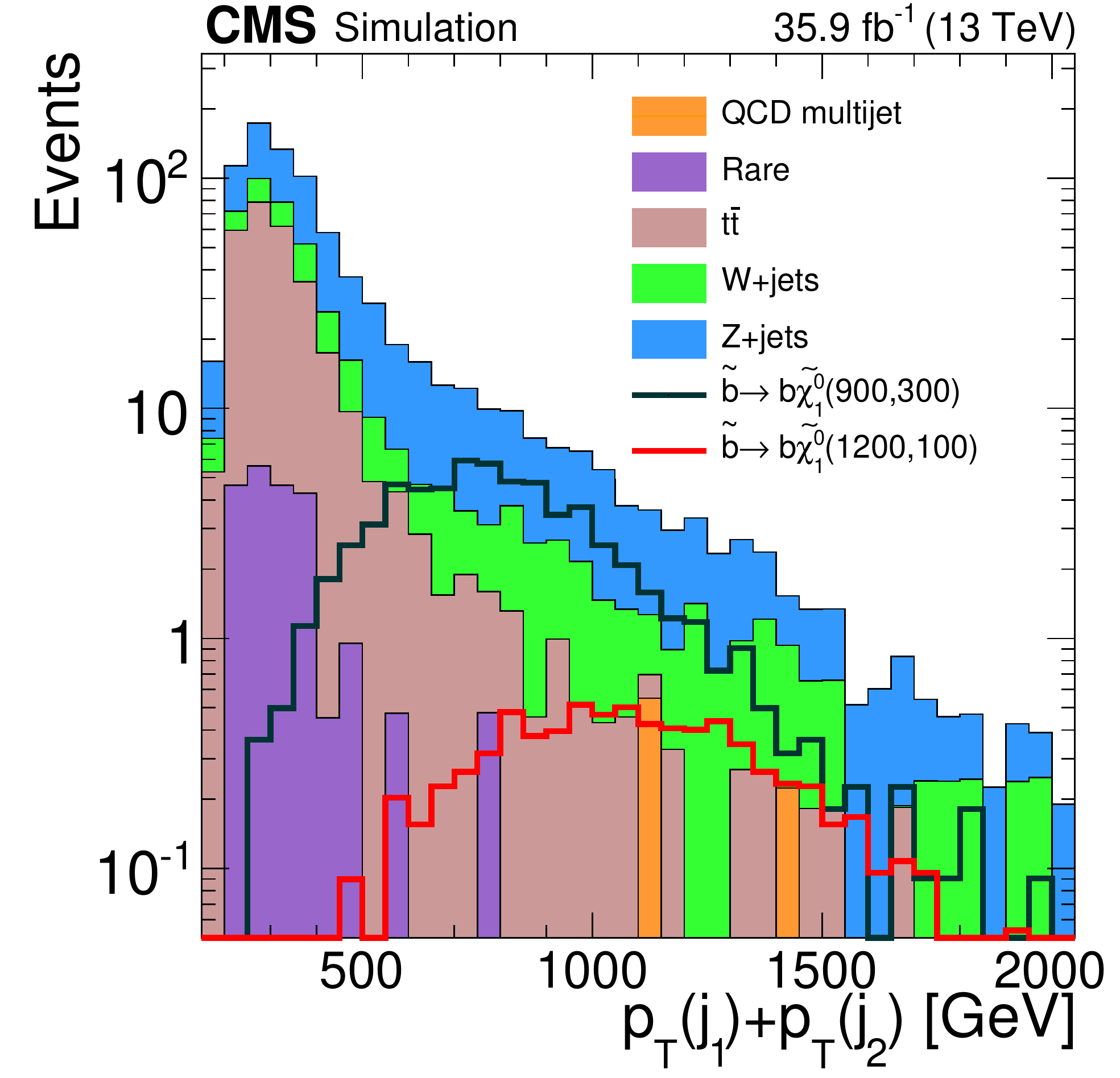} \\
\caption{Distribution of \mct (\cmsLeft) and $\pt(\text{j}_{1})+\pt(\text{j}_{2})$ (\cmsRight) for the searches in  noncompressed regions from simulation.
The stacked, filled histograms represent
different background components while the lines show two signal models with different bottom squark and neutralino mass hypotheses, ($m_{\sBot} = 900\GeV$ and $m_{\chiz_1} = 300\GeV$) and ($m_{\sBot} = 1200\GeV$ and $m_{\chiz_1} = 100\GeV$).}
\label{fig:noncom}
\end{figure}

\begin{figure}[!ht]
\centering
\includegraphics[width=\cmsFigWidth]{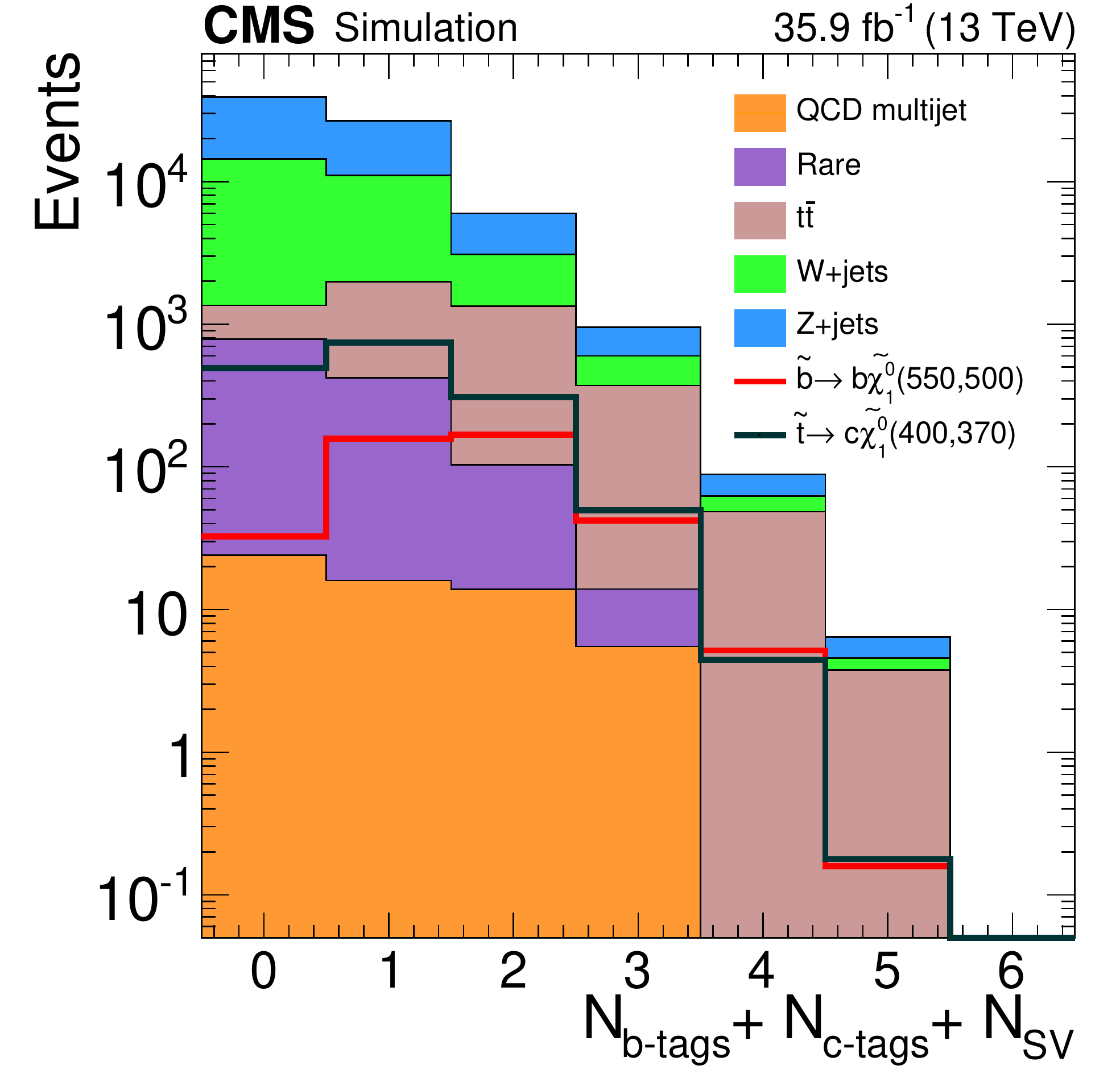}
\includegraphics[width=\cmsFigWidth]{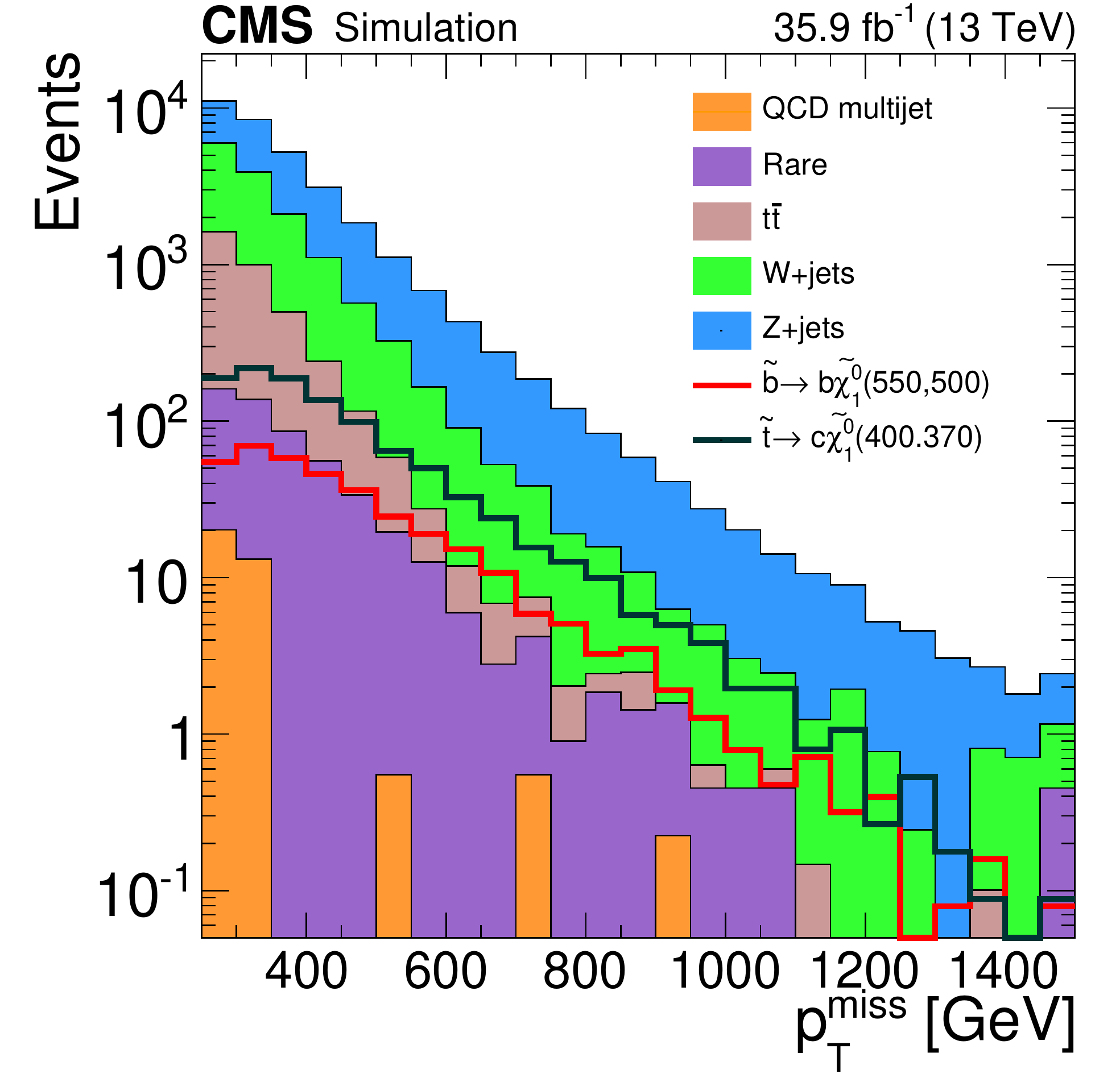} \\
\caption{Distributions of the combined \PQb-, \PQc-tagged jet, and SV multiplicity (\cmsLeft), and \met for events with at least one \PQb- or \PQc-tagged jet (\cmsRight), after the baseline selection for the compressed mass spectrum analysis, as obtained from simulation. The stacked, filled histograms represent different background components while the lines show two signal models with different bottom and top squark and neutralino mass hypotheses, ($m_{\sBot} = 550\GeV$ and $m_{\chiz_1} = 500\GeV$) and ($m_{\sTop} = 400\GeV$ and $m_{\chiz_1} = 370\GeV$).}
\label{fig:com}
\end{figure}

The discriminating power of the
kinematic quantities used in the analysis is shown in Figs.~\ref{fig:noncom} and~\ref{fig:com}. In the noncompressed region, the distributions of \mct and $\pt(\text{j}_{1})+\pt(\text{j}_{2})$, after applying all selection requirements (defined in Table~\ref{tab:sel}), are shown in Fig.~\ref{fig:noncom}. The combined number of \PQb-, \PQc-tagged jets and SV multiplicity for all events passing selection requirements in the compressed region is shown in the left panel of Fig.~\ref{fig:com}. The \met distribution for the events with at least one \PQb- or \PQc-tagged jet is shown in the right panel of Fig.~\ref{fig:com}.

\section{Background estimation}
\label{sec:backgrounds}

The SM background contributions originating from \znunu, \wjets, \ttbar, single-top-quark and QCD multijet processes are estimated from dedicated data control regions as discussed below. Smaller
contributions from other, rarer SM processes are estimated from simulation, and a conservative uncertainty of 50\% is assigned to these contributions~\cite{SUS16008}. In this paper the background from \wjets, \ttbar, and single top quark processes, is referred to as ``lost-lepton background".

\subsection{ \texorpdfstring{\znunu}{Z invisible} background estimation}
\label{sec:zinv_dy}

The \znunu background is estimated from a high-purity data sample of \zll~events in which we remove the leptons and recalculate the relevant kinematic variables to emulate \znunu\ events. The triggers used to collect this control sample require the presence of one or two muons or two electrons. For the single-muon trigger, the muon must have $ \pt > 50\GeV$; for the double muon (electron) triggers, the two highest-\pt muons (electrons) must have $ \pt > 17\GeV$ (23\GeV), and 8\GeV (12\GeV), respectively. The single muon trigger is used to recover a few percent efficiency loss that affects the double muon trigger in the high \pt muon region ($\pt > 400$\GeV). In keeping with the trigger constraints, the sample is selected by requiring the presence of two isolated leptons in the event with $|\eta|< 2.4$, and with $ \pt > 25$ or $> 20\GeV$ for the leading and subleading leptons, respectively. The invariant mass of the opposite-charge and same-flavour dilepton pair is required to be within 15\GeV of the \PZ~boson mass~\cite{PDG}. Each lepton is required to be separated from jets in the event by $\Delta R > 0.3$.

Apart from the lepton selection in the \zll control sample, the same object and event selection criteria, as described in
Section~\ref{sec:sel}, are applied to these events, which are subdivided into control regions, corresponding to the noncompressed and compressed search regions.

The expected number of \znunu events in each signal region is then
obtained by scaling the simulated yield, $N^{\text{MC}}_{\znunu}$,  by scale and shape correction
factors, according to:
\begin{equation}
N^{\text{Pred}}_{\znunu} =  N^{\text{MC}}_{\znunu} \frac{N^{\text{data}}_{\zll}}{N^{\text{MC}}_{\zll}}\, \mathrm{S}_{\mathrm{data/MC}}.
\label{equ:Zvv}
\end{equation}

The term $ N^{\text{data}}_{\zll}/N^{\text{MC}}_{\zll}$ is a scale factor to account for data-MC differences in the dilepton selection. It is computed for each \Nb, \Nc,
and \NI category separately, with an inclusive selection in the
kinematic variables \mct, \met, and \HT to improve statistical
precision. The term $\mathrm{S}_{\mathrm{data/MC}}$ is a shape correction factor that
accounts for possible differences in the shape of the kinematic variables used to define the signal regions. To compensate for the low event count due to the low branching fraction of the \PZ boson to dilepton final states, relaxed heavy flavor tagging requirements are used to compute the shape corrections. In the noncompressed region, jets are b tagged using a
loose working point, while in the compressed region an inclusive \Nb, \Nc, and \NI selection is used. The shape correction factors
in the noncompressed region are determined via comparison of the \mct distribution in  \zll events in simulation and data. To do the comparison, we first normalize the simulation to
the number of observed events in data after applying the loose selection criteria. The small contamination
from \ttbar, $\PW$+jets, single top quark and rare processes is estimated using simulation and subtracted
from data. The size of shape corrections in the noncompressed region varies between 3 to 20\% from lowest to highest \mct bin.
After applying the shape correction factor in bins of \mct and similar selections as in the search regions, good agreement between the data and simulation is
found as a function of \met and \HT. In the given \HT bin, the small residual difference in the \HT
distribution is considered as a systematic uncertainty.
In addition to the shape correction factors, the scale factor is calculated in the \zll control sample using the same b tagging requirements as in the signal region, and the value is determined to be consistent with unity within the statistical uncertainty.

For compressed regions, the shape correction factors are calculated inclusively in \Nb, \Nc, and \NI as a function of \met in the same way as in the noncompressed regions. The typical range of shape corrections in the compressed region is 5 to 70\%. The scale factors are determined in each \Nb, \Nc, and \NI signal region separately, and are consistent with unity within the statistical uncertainties.

Two sources of systematic uncertainty in the \znunu background contribution are uncertainties related to the use of simulation and uncertainties in the methods used to predict the background. The first set of uncertainties is related to the choice of the renormalization and factorization scales, PDFs, jet and \met energy scale, and the uncertainties in scale factors to correct the differences between the data and simulation in $\PQb$ or $\PQc$ tagging,  and lepton identification and isolation efficiencies. The total uncertainty from these sources is in the range of 1--20\%, depending on the signal region.

The second set of systematic uncertainties has a larger impact on the prediction, varies from 10 to 100\%, is due to the statistical uncertainties in the normalization and scale factors, contamination of other background sources in dilepton sample, the effect of the difference in
the \HT shape, and the uncertainty related to the trigger efficiency.

\subsection{Lost-lepton background estimation}
\label{sec:lostlep}

The lost-lepton background in each search region is estimated from a single-lepton control region in data selected by inverting
the muon or electron vetoes in the events collected with the same trigger as used to record the signal sample.
The control regions are defined through the same selection criteria as the corresponding search regions, including requirements on \HT, \mct, \Nb, \Nc, \NI, and \met, to remove any dependence of the prediction on the modelling of these kinematic variables in simulation. The possible contamination from signal in the single-lepton control region is found to have a negligible effect (${<}1\%$).
The lost-lepton component of the SM background in each search region, $N^\text{pred}_{\rm LL}$, is estimated from the corresponding data  via a transfer factor, $T_\mathrm{LL}$, determined from simulation:
\begin{equation}
\label{equ:lostlep}
N^\mathrm{Pred}_{\rm LL} = N^\text{data}_{1\ell} T_\mathrm{LL},\quad T_\mathrm{LL} = \frac{N^\mathrm{MC}_{0\ell}}{N^\mathrm{MC}_{1\ell}},
\end{equation}
where  $ N^\text{data}_{1\ell}$ is the observed event yield in the single-lepton control region and $ N^\mathrm{MC}_{0\ell}$ and $ N^\mathrm{MC}_{1\ell}$ are the simulated lost-lepton background yields in the corresponding
zero- and single-lepton regions, respectively. The transfer factor $T_\mathrm{LL}$ accounts for effects related to lepton acceptance and efficiency.

The largest uncertainty in the lost-lepton  background estimate is from statistical uncertainties in the event yields, ranging from 1 to 60\%, depending on the search region.
Contributions to the control regions from \zll~and rare processes are subtracted using estimates from simulation, where a 50\% uncertainty applied to the subtraction that leads to an uncertainty of 3--10\% in the lost-lepton  background prediction. The uncertainties related to discrepancies between the lepton selection efficiency in data and simulation give rise to a 3--4\% uncertainty in the final estimate. An additional uncertainty of 7\% in the \tauh component accounts for differences in isolation efficiency between muons and single-prong \tauh decays, as determined from studies with simulated samples of \wjets and \ttbar events.
A systematic uncertainty of 8--25\% is found for the uncertainties in \PQb or \PQc~tagging scale factors that are applied to the simulation for the differences in $\PQb$ or $\PQc$ tagging performance between data and simulation.

Finally, we estimate a systematic uncertainty in the transfer factor to account for differences in the \ttbar and \wjets composition of the search and control regions. This results in a 1--25\% uncertainty in the final prediction.

Table~\ref{tab:noncom-sys} provides a detailed breakdown of the various components
of the systematic uncertainties in the noncompressed and compressed regions.

\begin{table*}[!ht]
\centering
\topcaption{Different systematic
uncertainties in the lost-lepton background estimate.}
\label{tab:noncom-sys}
\ifthenelse{\boolean{cms@external}}{}{\resizebox{\textwidth}{!}}
{
\begin{tabular}{l| c | c}
\multicolumn{1}{c|}{Source} & Noncompressed regions (\%)& Compressed regions (\%) \\
\hline
b tagging efficiency  & 12--25 & \x8--22  \\
c tagging efficiency  & --- & 11--23\ \\
Lepton efficiency   & 3--4 & 3--4 \\
$\PGt_{\rm h}$ veto   & 7  & 7 \\
Transfer factor (statistical uncertainty)  & \x5--60 & \x1--40    \\
Transfer factor (systematic uncertainty)  & \x1--20 & 15--25    \\
Other SM process contamination  &  3--5 & \x3--10     \\
\end{tabular}
}
\end{table*}

\subsection{Multijet background estimation}
\label{sec:qcd}

The $\dphimet > 0.4$ requirement reduces the QCD multijet contribution to a small fraction of the total background in all search regions for both compressed and noncompressed models.  We estimate this contribution for each search region by applying a transfer factor to the number of events observed in control regions enriched in QCD events. The control regions are obtained by inverting the $\dphimet$ requirement. The transfer factor (\TQCD) is the ratio between the number of QCD multijet events in $\dphimet > 0.4$ to the number of events with $\dphimet < 0.4$, which is measured in simulation and validated with data in a sideband region with $\met \in[200,250]\GeV$ and similar selections as in the search regions. The estimated contribution from other SM processes (\ttbar,  \wjets, single top quark, and rare process production) based on simulated samples is subtracted from the event yields in the control region.

The transfer factor for the noncompressed regions does not vary significantly as a function of \HT or \mct. Therefore, we extract the value of \TQCD used for the noncompressed search regions from simulation and a low-\met sideband region selected with an inclusive requirement on \HT and \mct to reduce the statistical uncertainty in the transfer factor. The transfer factors for the compressed search regions are obtained from simulation and low-\met sidebands that are subdivided by the number of \PQb- and \PQc-tagged jets, and selected SV according to $\Nb+\Nc+\NI = 0$, $\Nb \geq 1$, $\Nc \geq 1$, and $\NI \geq 0$ regions. The $\Nb \geq 1$ ( $\Nc \geq 1$) regions are defined for extracting the QCD multijet background predictions for the $\Nb (\Nc) = 1$ and $\Nb (\Nc) = 2$ search regions.

The statistical uncertainties due to the limited number of events in the data control regions and the simulated samples are propagated to the final QCD multijet estimate, and range between 10 to 100\%.  The main uncertainty in \TQCD also originates from the statistical uncertainty of the observed and simulated event yields in the low-\met sideband region. We assign additional uncertainties for the differences in the \PQb and \PQc~tagging efficiencies between data and simulation.

\section{Results and interpretation}
\label{sec:result}

\begin{figure*}[!ht]
\centering
\includegraphics[width=\cmsFigWidth]{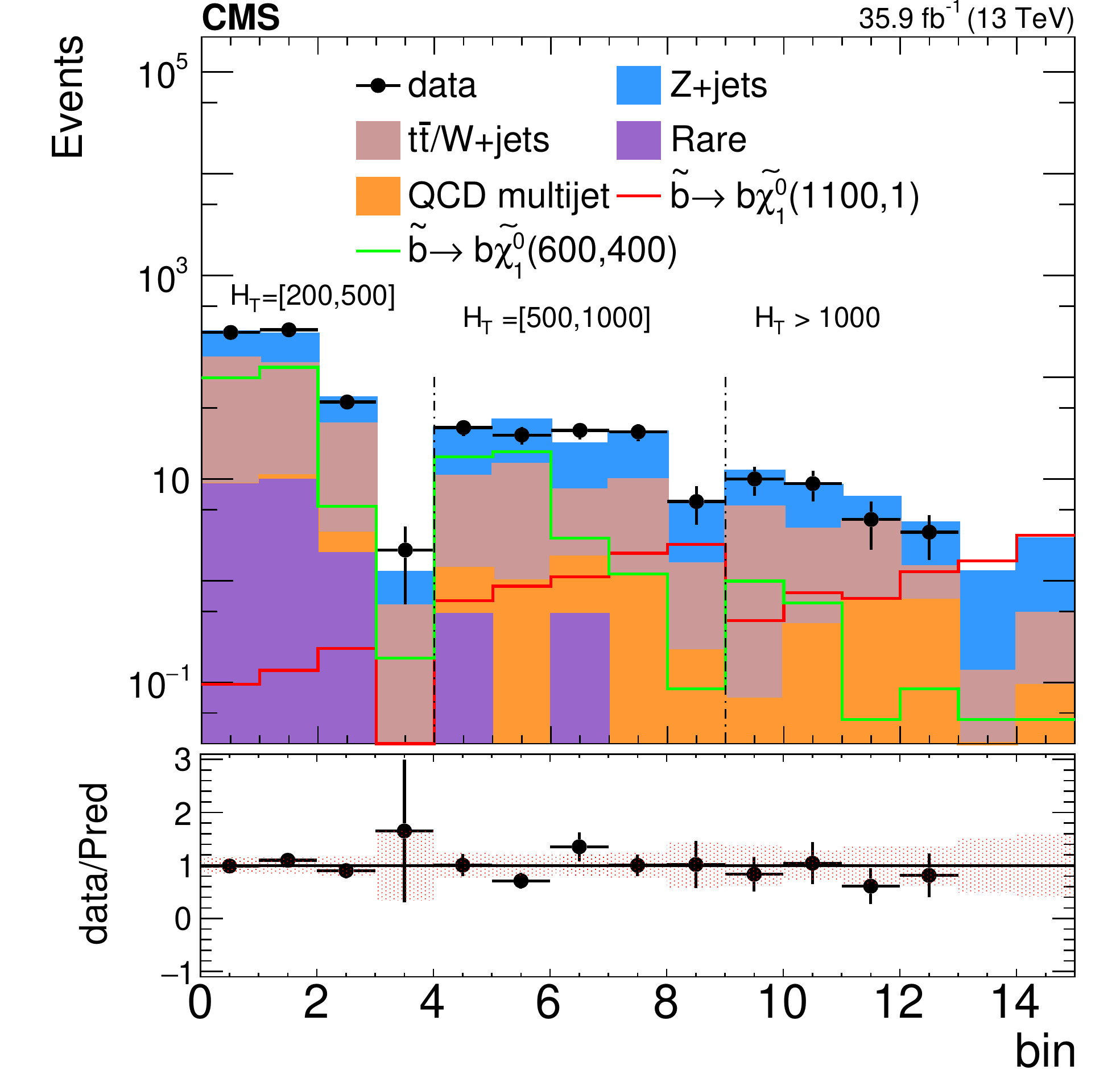}
\includegraphics[width=\cmsFigWidth]{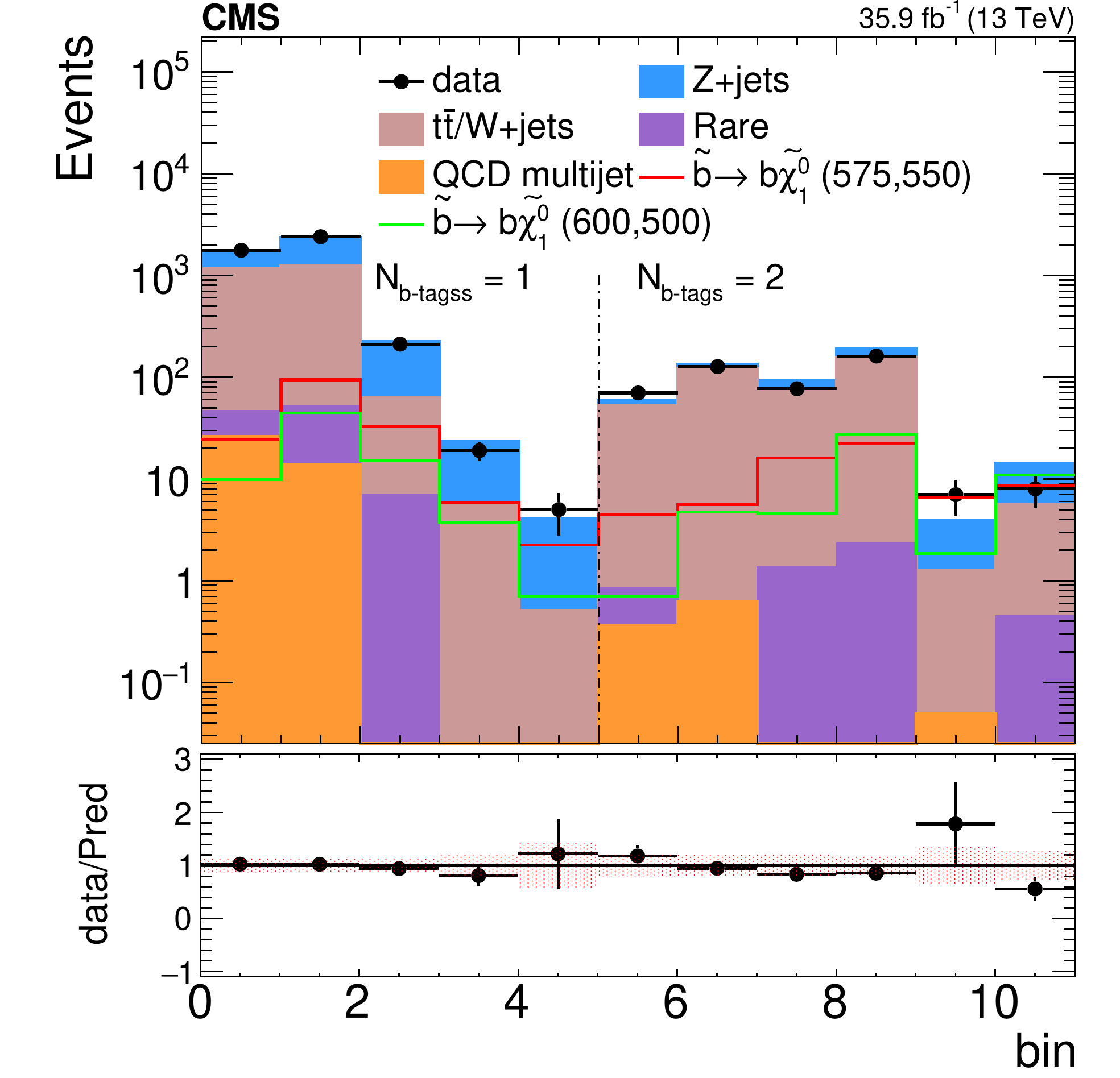} \\
\includegraphics[width=\cmsFigWidth]{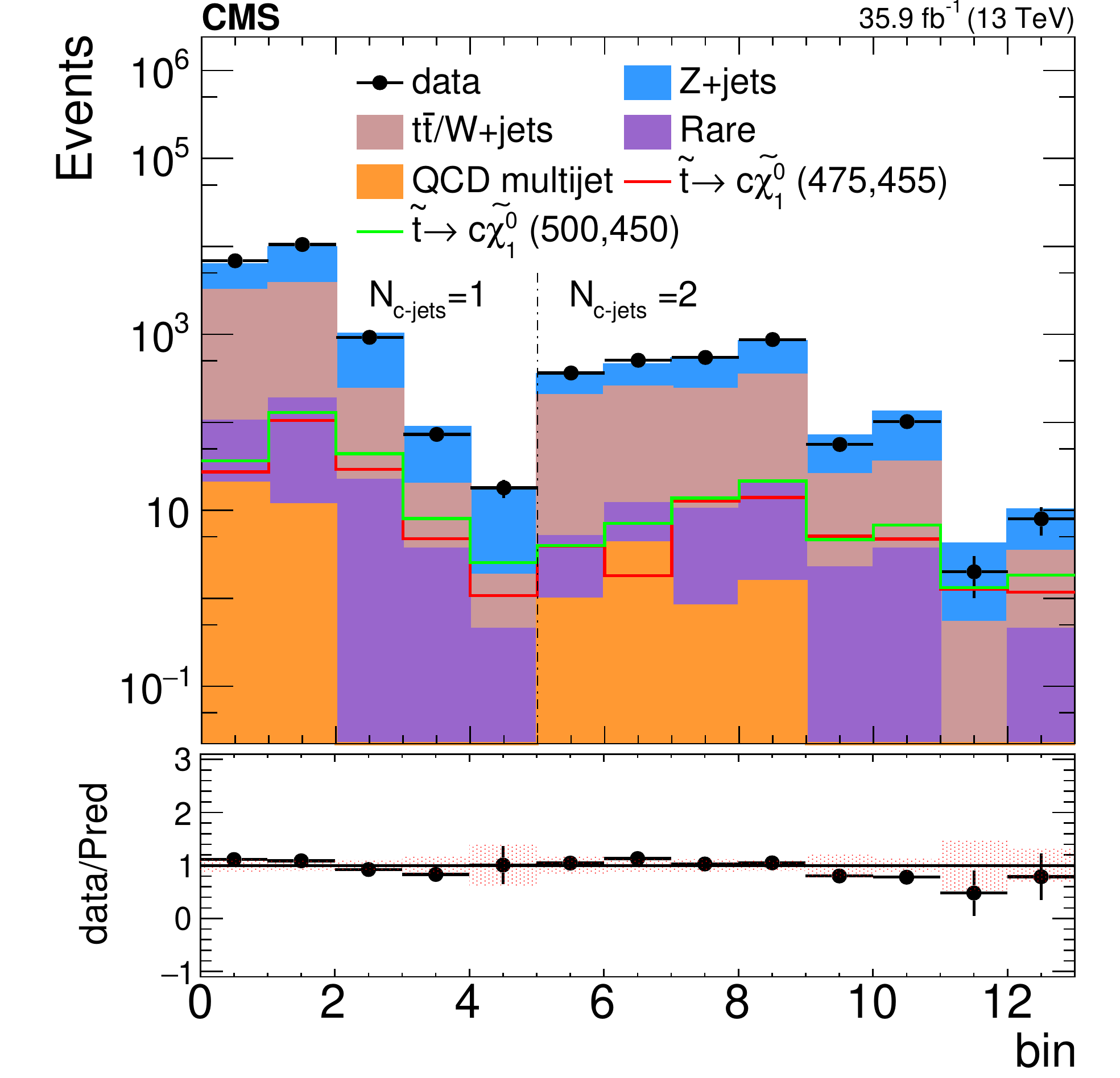}
\includegraphics[width=\cmsFigWidth]{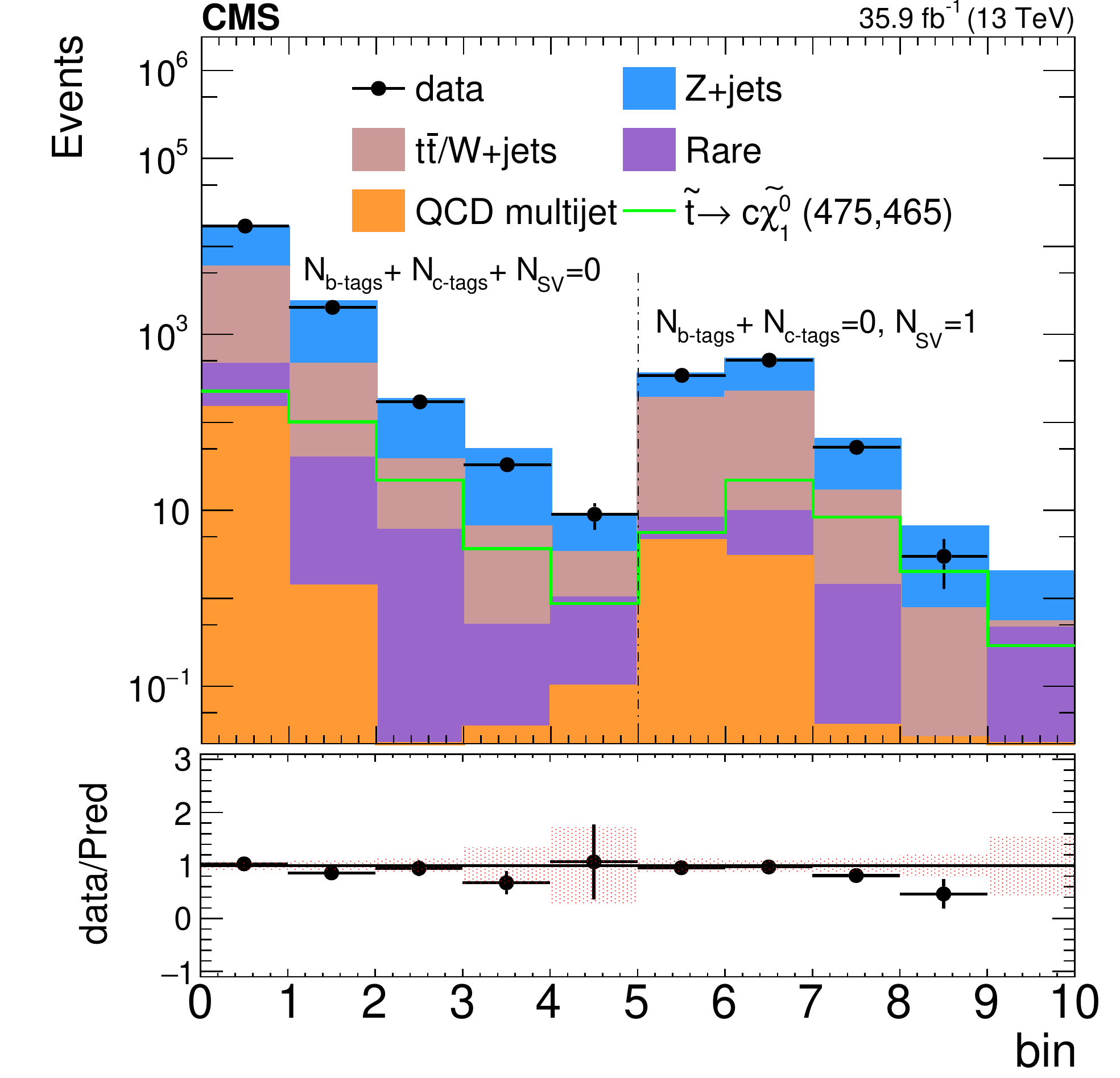}\\
\caption{Yields in the signal regions targeting the noncompressed (top left) and compressed (top
right: $\Nb = 1,\, 2$, bottom left: $\Nc = 1,\, 2$, bottom right: $\Nb+ \Nc = 0$) scenarios. Data are shown as black points. The background predictions are represented by the stacked, filled histograms. The expected yields for several signal models are also shown. The lower panels show the ratio of data over total background prediction in each signal region. The hatching indicates the total uncertainty in the background predictions.}
\label{fig:allPred}
\end{figure*}

The expected SM background yields and the number of events observed in data are
summarized in Table~\ref{tab:noncom_all} for the noncompressed search regions, and in Tables~\ref{tab:com_nb1},~\ref{tab:com_nc1}, and~\ref{tab:com_nI} for the compressed search regions. The results are shown in Fig.~\ref{fig:allPred} for both search regions.

\begin{table*}[!htp]
\renewcommand{\arraystretch}{1.1}
\topcaption{Observed number of events and background prediction in the noncompressed regions.
The total uncertainties in the background predictions are shown.}
\label{tab:noncom_all}
\centering
\ifthenelse{\boolean{cms@external}}{}{\resizebox{\textwidth}{!}}
{
\begin{tabular}{c| c c c c c c c c}
\multicolumn{9}{c}{Noncompressed regions}    \\ \hline
\HT~[\GeVns{}] &    \mct~[\GeVns{}] & Bin&  \znunu & Lost-lepton & QCD & Rare & Total SM & data    \\ \hline
\multirow{4}{*}{200--500}& 150--250 & 1&123$\pm$27&145$\pm$27&$<$0.7&8.8$\pm$4.4&278$\pm$40&275  \\
& 250--350 & 2&130$\pm$26&125$\pm$29&0.96$^{+1.67}_{-0.96}$&9.8$\pm$4.9&266$\pm$40&292 \\
& 350--450 & 3&28.5$\pm$9.1&31.6$\pm$7.2&1.06$^{+1.57}_{-1.06}$&1.87$\pm$0.93&\x63$\pm$12&57 \\
& $>$450 & 4&\x0.64$\pm$0.57&\x0.56$\pm$0.46&$<$0.30&$<$0.2&\x1.21$\pm$0.79&2  \\ \hline
\multirow{5}{*}{\x500--1000}& 150--250 & 5&21.2$\pm$6.6&\x9.2$\pm$3.7&0.85$^{+1.08}_{-0.85}$&0.47$\pm$0.24&31.8$\pm$7.6&32 \\
& 250--350 & 6&24.2$\pm$6.1&12.8$\pm$4.5&0.99$^{+1.3}_{-0.99}$&$<$0.2&37.9$\pm$7.8&27  \\
& 350--450 & 7&14.3$\pm$3.5&\x6.1$\pm$2.1&1.2$^{+1.6}_{-1.2}$&0.47$\pm$0.24&22.2$\pm$4.4&30 \\
& 450--600 & 8&19.1$\pm$6.2&\x8.6$\pm$2.3&1.1$^{+1.5}_{-1.1}$&$<$0.2&28.9$\pm$6.8&29 \\
& $>$600 &  9&\x4.4$\pm$2.4&\x1.25$\pm$0.67&$<$0.46&$<$0.2&\x5.7$\pm$2.5&6 \\  \hline
\multirow{6}{*}{${>}1000$}& 150--250 & 10&\x6.6$\pm$1.7&\x5.2$\pm$4.1&$<$0.23&$<$0.2&11.8$\pm$4.4&10\\
& 250--350 & 11&\x5.4$\pm$1.5&\x2.8$\pm$1.7&0.37$^{+0.53}_{-0.35}$&$<$0.2&\x8.6$\pm$2.3&9 \\
& 350--450 & 12&\x2.71$\pm$0.82&\x3.2$\pm$1.9&0.62$^{+0.80}_{-0.62}$&$<$0.2&\x6.6$\pm$2.3&4 \\
& 450--600 & 13&\x\x2.3$\pm$0.83&\x0.73$\pm$0.65&0.64$^{+0.82}_{-0.64}$&$<$0.2&\x3.7$\pm$1.3&3 \\
& 600--800 & 14&\x1.08$\pm$0.57&\x0.12$\pm$0.15&$<$0.13&$<$0.2&\x1.22$\pm$0.61&0 \\
& $>$800 &  15&\x2.1$\pm$1.4&\x0.38$\pm$0.40&$<$0.21&$<$0.2&\x2.5$\pm$1.5&0 \\
\end{tabular}
}
\end{table*}

\begin{table*}[!htp]
\renewcommand{\arraystretch}{1.1}
\topcaption{Observed number of events and the background prediction in the compressed regions with $\Nb = 1,\, 2$.
The total uncertainties in the background predictions are also shown.}
\label{tab:com_nb1}
\centering
\ifthenelse{\boolean{cms@external}}{}{\resizebox{\textwidth}{!}}
{
\begin{tabular}{c| c c c c c c c c}
\multicolumn{9}{c}{Compressed regions}    \\ \hline
\met~[\GeVns{}]& \HTb~[\GeVns{}]&Bin &    \znunu & Lost-lepton & QCD & Rare & Total SM & data   \\ \hline
\multicolumn{9}{c}{\Nb = 1}    \\ \hline
250--300&$<$100&1&555$\pm$92&1118$\pm$210&26$^{+27}_{-26}$&21$\pm$10&1720$\pm$230&1768 \\
300--500&$<$100&2&1100$\pm$130&1195$\pm$220&14$^{+15}_{-14}$&38$\pm$19&2348$\pm$260&2402 \\
500--750&$<$100&3&162$\pm$21&\x55$\pm$12&$<$0.33&6.7$\pm$3.5&224$\pm$25&211 \\
\x750--1000&$<$100&4&17.7$\pm$4.3&\x5.7$\pm$2.4&$<$0.15&$<$0.2&23.4$\pm$4.9&19  \\
$> 750$&$<$100&5&\x3.6$\pm$1.6&\x0.51$\pm$0.50&$<$0.1\x & $<$0.2 & \x4.1$\pm$1.7&5 \\ \hline
\multicolumn{9}{c}{\Nb = 2}    \\ \hline
250--300&$<$100&6&\x6.9$\pm$2.8&\x51$\pm$12&0.36$^{+0.46}_{-0.36}$&0.47$\pm$0.23&\x59$\pm$12&70 \\
250--300&100--200&7&12.9$\pm$4.5&120$\pm$25&0.62$^{+0.78}_{-0.62}$&$<$0.2&134$\pm$25&127 \\
300--500&$<$100&8&19.4$\pm$6.3&\x72$\pm$17&$<$0.2\x&1.36$\pm$0.68&\x92$\pm$18&77 \\
300--500& 100--200&9&\x34$\pm$10&151$\pm$31&$<$0.2\x&1.35$\pm$0.67&188$\pm$32&161 \\
${>}500$& $<$100&10&\x2.64$\pm$0.98&\x1.22$\pm$0.87&$<$0.1\x&$<$0.2&\x3.9$\pm$1.3&7 \\
${>}500$& 100--200&11&\x8.7$\pm$2.9&\x5.1$\pm$2.3&$<$0.1\x&0.45$\pm$0.22&14.35$\pm$3.7\x&8 \\
\end{tabular}
}
\end{table*}

\begin{table*}[!htp]
\renewcommand{\arraystretch}{1.1}
\topcaption{Observed number of events and the background prediction in the compressed regions with $\Nc = 1,\, 2$.
The total uncertainties in the background predictions are also shown. }
\label{tab:com_nc1}
\centering
\ifthenelse{\boolean{cms@external}}{}{\resizebox{\textwidth}{!}}
{
\begin{tabular}{c| c c c c c c c c }
\multicolumn{9}{c}{Compressed regions}    \\ \hline
\met~[\GeVns{}]& \HTc~[\GeVns{}] &   Bin &    \znunu & Lost-lepton & QCD & Rare & Total SM & data   \\ \hline
\multicolumn{8}{c}{\Nc = 1}    \\ \hline
250--300&$<$100 &1&3022$\pm$480&3049$\pm$530&20$^{+22}_{-20}$&\x85$\pm$42&6177$\pm$720&6867 \\
300--500&$<$100&2&5852$\pm$690&3622$\pm$620&11$^{+12}_{-11}$&178$\pm$89&9664$\pm$930&10515\\
500--750 &$<$100&3&765$\pm$95&214$\pm$39&$<$0.2&\x22$\pm$11&1002$\pm$100&926 \\
\x750--1000&$<$100&4&\x67$\pm$13&16.2$\pm$3.9&$<$0.1&\x3.7$\pm$1.8&\x88$\pm$14&73 \\
${>}1000$&$<$100&5& 16.0$\pm$6.9&\x1.37$\pm$0.78&$<$0.1&\x0.45$\pm$0.22&17.8$\pm$7.1&18 \\ \hline
\multicolumn{9}{c}{\Nc = 2}    \\ \hline
250--300 & $<$100 &6&145$\pm$33&198$\pm$42&0.98$^{+1.1}_{-0.98}$&\x4.1$\pm$2.1&348$\pm$54&364 \\
250--300& 100--200 &7&199$\pm$25&238$\pm$46&4.3$\pm$4.7&\x7.8$\pm$3.9&449$\pm$53&508 \\
300--500&$<$100 &8&293$\pm$39&229$\pm$45&0.81$\pm$0.91&\x9.7$\pm$4.8&532$\pm$60&547 \\
300--500& 100--200 &9&489$\pm$55&323$\pm$59&1.5$\pm$1.7&19.3$\pm$9.6&833$\pm$81&874 \\
500--750&$<$100 &10&\x44$\pm$13&23.4$\pm$7.2&$<$0.1&\x2.3$\pm$1.1&\x70$\pm$15&56 \\
500--750& 100--200 &11&\x95$\pm$14&31.8$\pm$7.8&$<$0.1&\x3.7$\pm$1.8&130$\pm$16&102 \\
${>}750$&$<$100 &12&\x3.6$\pm$1.9&\x0.52$\pm$0.58&$<$0.1&$<$0.2&\x4.1$\pm$1.9&2 \\
${>}750$&100--200 &13&\x6.7$\pm$2.6&\x2.9$\pm$1.6&$<$0.1&\x0.45$\pm$0.22&10.1$\pm$3.1&8 \\
\end{tabular}
}
\end{table*}

\begin{table*}[!htp]
\renewcommand{\arraystretch}{1.1}
\topcaption{Observed number of events and the background prediction in the compressed regions with $\Nb+ \Nc = 0$.
The total uncertainties in the background predictions are also shown.}
\label{tab:com_nI}
\centering
\ifthenelse{\boolean{cms@external}}{}{\resizebox{\textwidth}{!}}
{
\begin{tabular}{c| c c c c c c c }
\multicolumn{8}{c}{Compressed regions}    \\ \hline
\met~[\GeVns{}]&   Bin &    \znunu & Lost-lepton & QCD & Rare & Total SM & data   \\  \hline
\multicolumn{8}{c}{$\Nb+ \Nc+ \NI = 0$}    \\ \hline
300--500&1&10676$\pm$740\x&5398$\pm$930&148$^{+160}_{-150}$&320$\pm$160&16542$\pm$1200&17042 \\
500--750&2&1902$\pm$180&414$\pm$73&1.4$^{+2.1}_{-1.4}$&39$\pm$19&2358$\pm$200&2028 \\
\x750--1000&3&143$\pm$21&31.2$\pm$6.6&$<$0.45&6.1$\pm$3.1&181$\pm$22&171 \\
1000--1250&4&\x42$\pm$16&\x5.9$\pm$2.8&$<$0.03&0.47$\pm$0.23&\x49$\pm$16&33 \\
${>}1250$&5&\x5.1$\pm$5.7&\x2.3$\pm$1.6&0.09$^{+0.17}_{-0.09}$&0.92$\pm$0.46&\x8.4$\pm$6.0&9 \\ \hline
\multicolumn{8}{c}{$\Nb+ \Nc = 0$, $\NI> 0$}    \\ \hline
250--300&6&169$\pm$22&179$\pm$36&4.5$^{+5.1}_{-4.5}$&3.7$\pm$1.9&357$\pm$43&331 \\
300--500&7&303$\pm$37&210$\pm$41&2.9$^{+3.3}_{-2.9}$&6.9$\pm$3.4&523$\pm$57&509 \\
500--750&8&46.6$\pm$6.2&15.1$\pm$4.8&0.03$^{+0.13}_{-0.03}$&1.40$\pm$0.70&64.2$\pm$7.8&52\\
\x750--1000&9&\x5.7$\pm$1.2&\x0.73$\pm$0.59&$<$0.1\x&$<$0.2&\x6.5$\pm$1.3&3 \\
${>}1000$&10&\x1.5$\pm$1.1&\x0.07$\pm$0.10&$<$0.2\x&0.45$\pm$0.22&\x2.0$\pm$1.1&0 \\
\end{tabular}
}
\end{table*}

The data are consistent with the background expected from the SM processes.
The results are interpreted as upper cross section limits on bottom and top squark pair production.

The dominant systematic uncertainties on the signal yield predictions are:
the luminosity determination (2.5\%)~\cite{CMS-PAS-LUM-17-001}, the signal
acceptance and efficiency arising from the jet energy corrections (5\%);
renormalization and factorization scale (5\%); ISR modelling (5--20\%); trigger efficiency (2\%); b and c tagging efficiency (5--30\%); and selected SV efficiency (16--50\%). The uncertainty of 16\% is considered if the selected SV is matched to  b hadrons, and it is doubled if the selected SV is matched to c hadrons. Finally, a 50\% uncertainty in the selected SV efficiency is applied, if it is not matched to either b or c hadrons. However, due to the small misidentification rate (~1\%) the considered 50\% uncertainty has a negligible effect on final limits. The statistical uncertainty due to the limited size of the simulated samples, calculated for each signal model, varies from a few percent to 100\% and is not correlated with signal systematic uncertainties. While the uncertainties in the \PQb- and \PQc-tagged jet and lepton efficiency corrections in simulation are correlated between
different processes and search bins, the uncertainties in transfer factors are
treated as fully uncorrelated. For the signal, all systematic uncertainties are correlated between the different search regions.
We improve the modeling of ISR jets, which affects the total transverse momentum (\pt(ISR)) of the system of SUSY particles, by reweighting the \pt(ISR) distribution of signal events. This reweighting procedure is based on studies of the transverse momentum of \PZ events~\cite{CMS-STOP-lepton}. The reweighting factors range between 1.18 at \pt(ISR)~125 GeV and 0.78 for $\pt(ISR) > 600$\GeV. We take the deviation from 1.0 as the systematic uncertainty in the reweighting procedure.

The 49 signal bins in \met, \HT, \mct, \Nb, \Nc, and \NI are statistically
independent, and the
correlations among all the systematic uncertainties in different bins are taken into account.
The 95\% confidence level (CL) upper limits on SUSY production cross-sections are calculated using a modified frequentist approach with the CL$_\mathrm{S}$ criterion~\cite{CLs1,CLs,combine} in which a profile likelihood rate test-statistic is used. The limits are determined using asymptotic approximations for the distributions of the test-statistic~\cite{asy}.

\begin{figure}[!ht]
\centering
\includegraphics[width=0.48\textwidth]{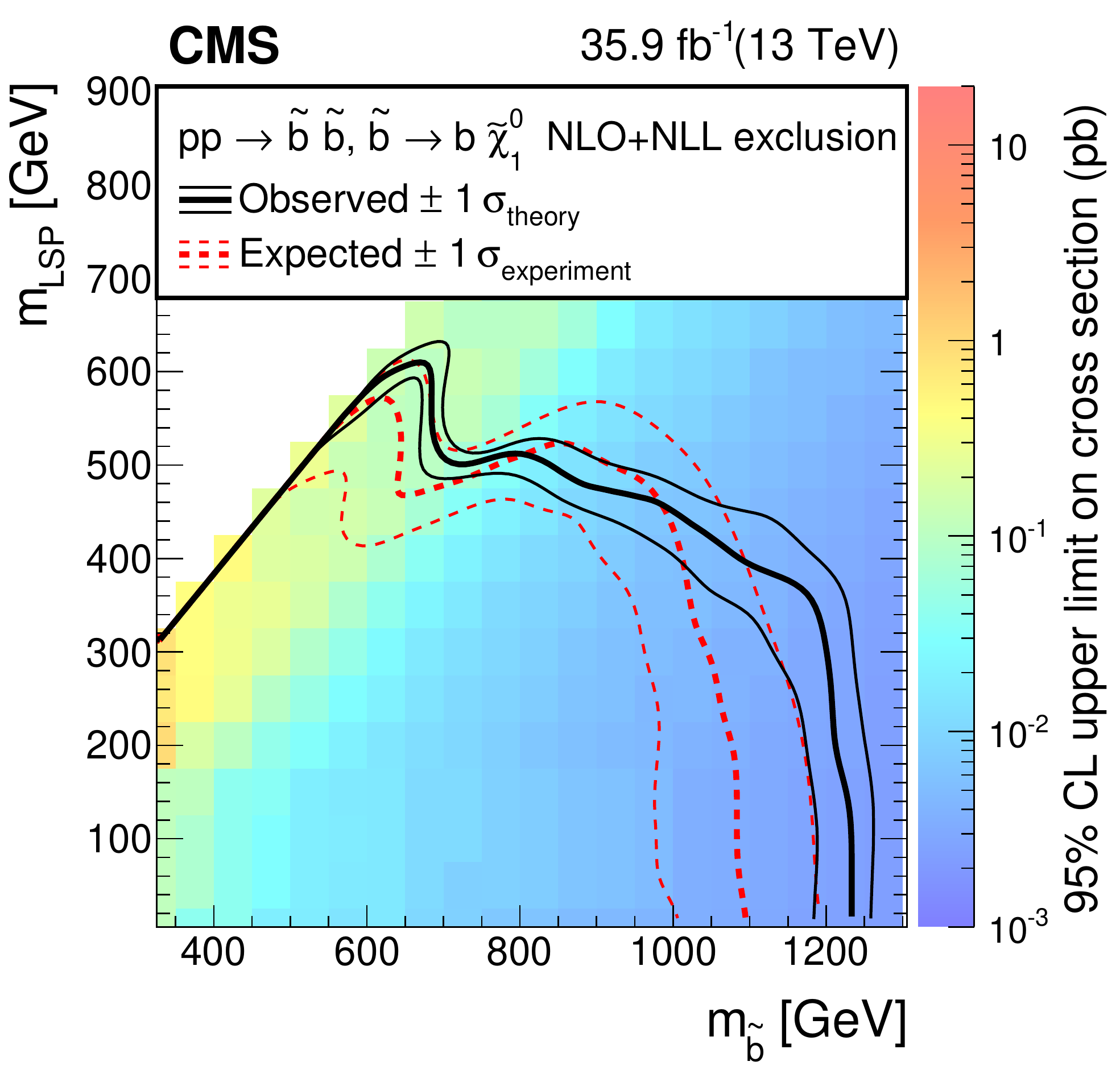}
\caption{Exclusion limits at 95\% CL for direct bottom squark pair production for the decay mode $\sbottomq\to\PQb \lsp$.
The regions enclosed by the black curves represent the observed exclusion and the ${\pm}1$ standard deviation for the NLO+NLL cross section calculations and their uncertainties~\cite{Borschensky:2014cia}. The dashed red lines indicate the expected limits at 95\% CL and their ${\pm}1$ standard deviation experimental uncertainties. }
\label{fig:Exp1}
\end{figure}

Figure \ref{fig:Exp1} shows the expected and observed 95\% CL upper limits on the bottom squark cross sections, assuming the bottom squark exclusively decays to a bottom quark and an LSP.

\begin{figure}[!ht]
\centering
\includegraphics[width=0.48\textwidth]{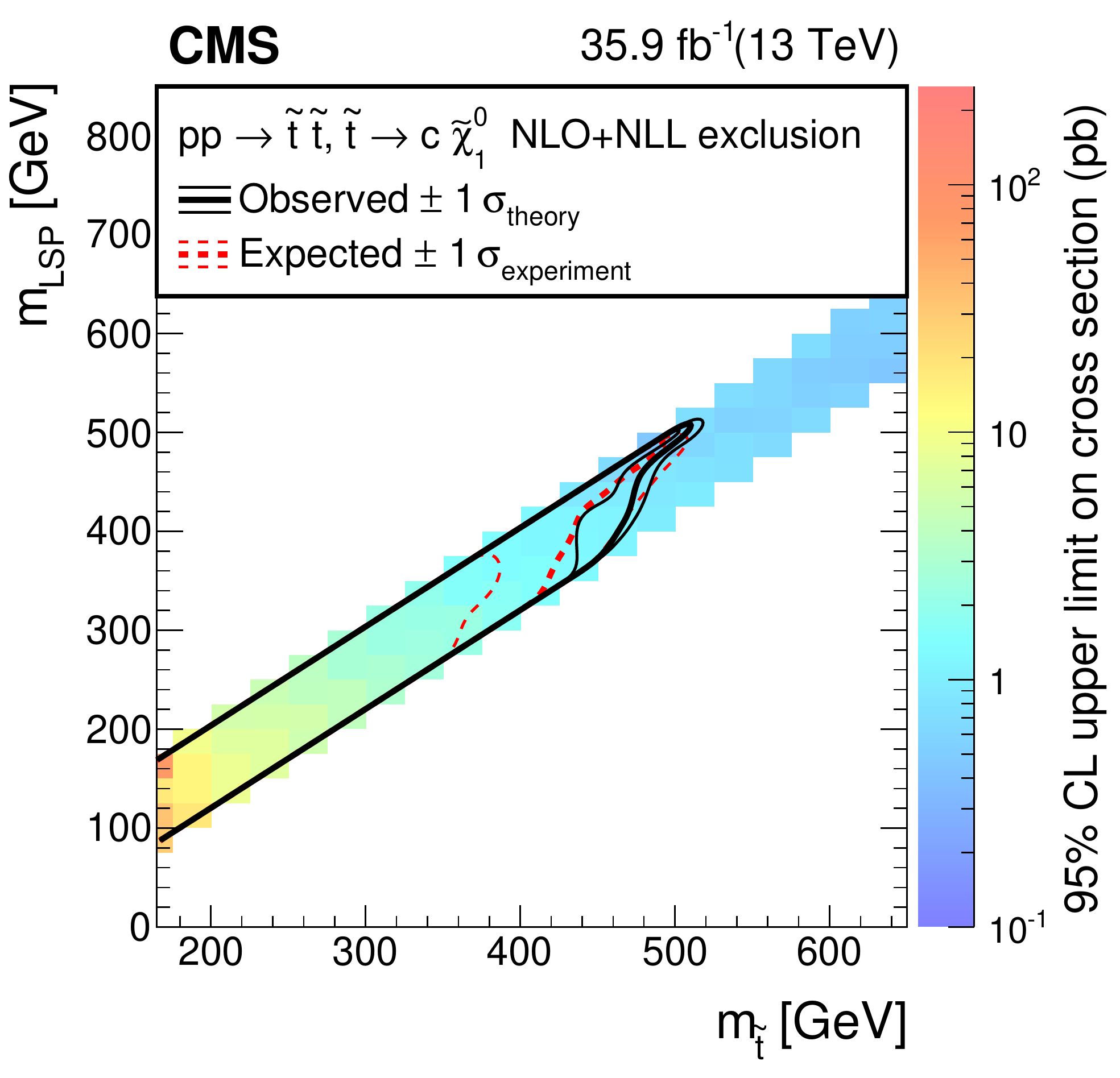}
\caption{The combined 95\% CL exclusion limits for top squark pair production assuming 100\% branching fraction to the decay
$\sTop\to \PQc \chiz_1$.
Notations are as in Fig~\ref{fig:Exp1}.}
\label{fig:Exp2}
\end{figure}

Both compressed and noncompressed regions are used to search for the bottom squark, and the compressed search regions are only used to set upper limits on the top squark cross sections when the mass splitting between the top squark and the LSP is smaller than the mass of the \PW~boson. Figure~\ref{fig:Exp2} shows the expected and observed 95\% CL upper limits on the top squark cross sections in the $m_{\stopq}$-$m_{\lsp}$ plane assuming the top squark decays exclusively to a charm quark and an LSP. Top squarks with masses below 510\GeV are excluded in this model for a mass splitting between the top squark and the LSP is small. For the similar interpretation in~\cite{CMS-SUS-16-049}, top squark and LSP masses are excluded up to 560 and 520\GeV, respectively.

To facilitate reinterpretation, the covariance matrices for the background estimates in the compressed and noncompressed search regions are provided in supplemental \ifthenelse{\boolean{cms@external}}{}{Appendix~}\ref{sec:app_result}.

\section{Summary}
\label{sec:con}

A search for the pair production of third-generation squarks is performed using data collected by the CMS experiment, focusing on two-body decays to bottom or charm quarks. For bottom-squark pair production, the decay mode considered is $\sbottomq\to\PQb\lsp$, while for top-squark pair production, the decay mode considered is $\stopq\to\PQc\lsp$, a flavor-changing neutral current process. No statistically significant excess of events is observed above the expected standard model background, and exclusion limits are set at 95\% confidence level in the context of simplified models of direct top and bottom squark pair production. Bottom squark masses below 1220\GeV are excluded assuming that the lightest supersymmetric particle (LSP) is massless; bottom squark masses below 675\GeV are excluded for LSP masses up to 600\GeV.
Top squark masses below 510\GeV are excluded for the scenario in which $\stopq\to\PQc\lsp$ and the mass splitting between the top squark and the LSP is small.

\begin{acknowledgments}

\hyphenation{Bundes-ministerium Forschungs-gemeinschaft Forschungs-zentren Rachada-pisek} We congratulate our colleagues in the CERN accelerator departments for the excellent performance of the LHC and thank the technical and administrative staffs at CERN and at other CMS institutes for their contributions to the success of the CMS effort. In addition, we gratefully acknowledge the computing centres and personnel of the Worldwide LHC Computing Grid for delivering so effectively the computing infrastructure essential to our analyses. Finally, we acknowledge the enduring support for the construction and operation of the LHC and the CMS detector provided by the following funding agencies: the Austrian Federal Ministry of Science, Research and Economy and the Austrian Science Fund; the Belgian Fonds de la Recherche Scientifique, and Fonds voor Wetenschappelijk Onderzoek; the Brazilian Funding Agencies (CNPq, CAPES, FAPERJ, and FAPESP); the Bulgarian Ministry of Education and Science; CERN; the Chinese Academy of Sciences, Ministry of Science and Technology, and National Natural Science Foundation of China; the Colombian Funding Agency (COLCIENCIAS); the Croatian Ministry of Science, Education and Sport, and the Croatian Science Foundation; the Research Promotion Foundation, Cyprus; the Secretariat for Higher Education, Science, Technology and Innovation, Ecuador; the Ministry of Education and Research, Estonian Research Council via IUT23-4 and IUT23-6 and European Regional Development Fund, Estonia; the Academy of Finland, Finnish Ministry of Education and Culture, and Helsinki Institute of Physics; the Institut National de Physique Nucl\'eaire et de Physique des Particules~/~CNRS, and Commissariat \`a l'\'Energie Atomique et aux \'Energies Alternatives~/~CEA, France; the Bundesministerium f\"ur Bildung und Forschung, Deutsche Forschungsgemeinschaft, and Helmholtz-Gemeinschaft Deutscher Forschungszentren, Germany; the General Secretariat for Research and Technology, Greece; the National Scientific Research Foundation, and National Innovation Office, Hungary; the Department of Atomic Energy and the Department of Science and Technology, India; the Institute for Studies in Theoretical Physics and Mathematics, Iran; the Science Foundation, Ireland; the Istituto Nazionale di Fisica Nucleare, Italy; the Ministry of Science, ICT and Future Planning, and National Research Foundation (NRF), Republic of Korea; the Lithuanian Academy of Sciences; the Ministry of Education, and University of Malaya (Malaysia); the Mexican Funding Agencies (BUAP, CINVESTAV, CONACYT, LNS, SEP, and UASLP-FAI); the Ministry of Business, Innovation and Employment, New Zealand; the Pakistan Atomic Energy Commission; the Ministry of Science and Higher Education and the National Science Centre, Poland; the Funda\c{c}\~ao para a Ci\^encia e a Tecnologia, Portugal; JINR, Dubna; the Ministry of Education and Science of the Russian Federation, the Federal Agency of Atomic Energy of the Russian Federation, Russian Academy of Sciences, the Russian Foundation for Basic Research and the Russian Competitiveness Program of NRNU ``MEPhI"; the Ministry of Education, Science and Technological Development of Serbia; the Secretar\'{\i}a de Estado de Investigaci\'on, Desarrollo e Innovaci\'on, Programa Consolider-Ingenio 2010, Plan de Ciencia, Tecnolog\'{i}a e Innovaci\'on 2013-2017 del Principado de Asturias and Fondo Europeo de Desarrollo Regional, Spain; the Swiss Funding Agencies (ETH Board, ETH Zurich, PSI, SNF, UniZH, Canton Zurich, and SER); the Ministry of Science and Technology, Taipei; the Thailand Center of Excellence in Physics, the Institute for the Promotion of Teaching Science and Technology of Thailand, Special Task Force for Activating Research and the National Science and Technology Development Agency of Thailand; the Scientific and Technical Research Council of Turkey, and Turkish Atomic Energy Authority; the National Academy of Sciences of Ukraine, and State Fund for Fundamental Researches, Ukraine; the Science and Technology Facilities Council, UK; the US Department of Energy, and the US National Science Foundation.

Individuals have received support from the Marie-Curie programme and the European Research Council and Horizon 2020 Grant, contract No. 675440 (European Union); the Leventis Foundation; the A. P. Sloan Foundation; the Alexander von Humboldt Foundation; the Belgian Federal Science Policy Office; the Fonds pour la Formation \`a la Recherche dans l'Industrie et dans l'Agriculture (FRIA-Belgium); the Agentschap voor Innovatie door Wetenschap en Technologie (IWT-Belgium); the Ministry of Education, Youth and Sports (MEYS) of the Czech Republic; the Council of Scientific and Industrial Research, India; the HOMING PLUS programme of the Foundation for Polish Science, cofinanced from European Union, Regional Development Fund, the Mobility Plus programme of the Ministry of Science and Higher Education, the National Science Center (Poland), contracts Harmonia 2014/14/M/ST2/00428, Opus 2014/13/B/ST2/02543, 2014/15/B/ST2/03998, and 2015/19/B/ST2/02861, Sonata-bis 2012/07/E/ST2/01406; the National Priorities Research Program by Qatar National Research Fund; the Programa Clar\'in-COFUND del Principado de Asturias; the Thalis and Aristeia programmes cofinanced by EU-ESF and the Greek NSRF; the Rachadapisek Sompot Fund for Postdoctoral Fellowship, Chulalongkorn University and the Chulalongkorn Academic into Its 2nd Century Project Advancement Project (Thailand); and the Welch Foundation, contract C-1845.

\end{acknowledgments}

\clearpage
\newpage
\bibliography{auto_generated}

\providecommand{\href}[2]{#2}\begingroup\raggedright\begin{thebibliography}{10}%
\makeatletter
\providecommand{\hrefCMSnoop }[0]{\@secondoftwo}%
\makeatother
\providecommand{\doi}{\texttt{doi:}\begingroup \urlstyle{tt}\Url}

\bibitem{Barbi}
\hrefCMSnoop {}{R.~Barbieri and G.~F. Giudice, ``{Upper bounds on
  supersymmetric particle masses}'',} \textit{ Nucl. Phys. B} \textbf{ 306}
  (1988) 63,
\href{http://dx.doi.org/10.1016/0550-3213(88)90171-X}{\doi{10.1016/0550-3213(88)90171-X}}.
%%CITATION = NUPHA,B306,63;%%.

\bibitem{Ramond}
\hrefCMSnoop {}{P.~Ramond, ``{Dual Theory for Free Fermions}'',} \textit{ Phys.
  Rev. D} \textbf{ 3} (1971) 2415,
\href{http://dx.doi.org/10.1103/PhysRevD.3.2415}{\doi{10.1103/PhysRevD.3.2415}}.
%%CITATION = PHRVA,D3,2415;%%.

\bibitem{Golfand}
\href {http://www.jetpletters.ac.ru/ps/1584/article_24309.pdf}{{\relax Yu}.~A.
  Gol'fand and E.~P. Likhtman, ``Extension of the algebra of {P}oincar{\' e}
  group generators and violation of {P} invariance'',} \textit{ JETP Lett.}
  \textbf{ 13} (1971)
323.
%%CITATION = JTPLA,13,323;%%.

\bibitem{Neveu}
\hrefCMSnoop {}{A.~Neveu and J.~H. Schwarz, ``{Factorizable dual model of
  pions}'',} \textit{ Nucl. Phys. B} \textbf{ 31} (1971) 86,
\href{http://dx.doi.org/10.1016/0550-3213(71)90448-2}{\doi{10.1016/0550-3213(71)90448-2}}.
%%CITATION = NUPHA,B31,86;%%.

\bibitem{Volkov}
\href {http://www.jetpletters.ac.ru/ps/1766/article_26864.pdf}{D.~V. Volkov and
  V.~P. Akulov, ``{Possible universal neutrino interaction}'',} \textit{ JETP
  Lett.} \textbf{ 16} (1972) 438.
[Pisma Zh. Eksp. Teor. Fiz.16,621(1972)].
%%CITATION = JTPLA,16,438;%%.

\bibitem{Wess}
\hrefCMSnoop {}{J.~Wess and B.~Zumino, ``{A lagrangian model invariant under
  supergauge transformations}'',} \textit{ Phys. Lett. B} \textbf{ 49} (1974)
  52,
\href{http://dx.doi.org/10.1016/0370-2693(74)90578-4}{\doi{10.1016/0370-2693(74)90578-4}}.
%%CITATION = PHLTA,49B,52;%%.

\bibitem{Wess1}
\hrefCMSnoop {}{J.~Wess and B.~Zumino, ``{Supergauge transformations in four
  dimensions}'',} \textit{ Nucl. Phys. B} \textbf{ 70} (1974) 39,
\href{http://dx.doi.org/10.1016/0550-3213(74)90355-1}{\doi{10.1016/0550-3213(74)90355-1}}.
%%CITATION = NUPHA,B70,39;%%.

\bibitem{Fayet}
\hrefCMSnoop {}{P.~Fayet, ``{Supergauge invariant extension of the Higgs
  mechanism and a model for the electron and its neutrino}'',} \textit{ Nucl.
  Phys. B} \textbf{ 90} (1975) 104,
\href{http://dx.doi.org/10.1016/0550-3213(75)90636-7}{\doi{10.1016/0550-3213(75)90636-7}}.
%%CITATION = NUPHA,B90,104;%%.

\bibitem{Nilles}
\hrefCMSnoop {}{H.~P. Nilles, ``{Supersymmetry, supergravity and particle
  physics}'',} \textit{ Phys. Rept.} \textbf{ 110} (1984) 1,
\href{http://dx.doi.org/10.1016/0370-1573(84)90008-5}{\doi{10.1016/0370-1573(84)90008-5}}.
%%CITATION = PRPLC,110,1;%%.

\bibitem{Pap}
\hrefCMSnoop {}{M.~Papucci, J.~T. Ruderman, and A.~Weiler, ``{Natural SUSY
  endures}'',} \textit{ JHEP} \textbf{ 09} (2012) 035,
  \href{http://dx.doi.org/10.1007/JHEP09(2012)035}{\doi{10.1007/JHEP09(2012)035}},
\href{http://www.arXiv.org/abs/1110.6926}{\texttt{arXiv:1110.6926}}.
%%CITATION = ARXIV:1110.6926;%%.

\bibitem{Farrar:1978xj}
\hrefCMSnoop {}{G.~R. Farrar and P.~Fayet, ``{Phenomenology of the production,
  decay, and detection of new hadronic states associated with
  supersymmetry}'',} \textit{ Phys. Lett. B} \textbf{ 76} (1978) 575,
\href{http://dx.doi.org/10.1016/0370-2693(78)90858-4}{\doi{10.1016/0370-2693(78)90858-4}}.
%%CITATION = PHLTA,B76,575;%%.

\bibitem{darkmatter}
\hrefCMSnoop {}{G.~Jungman, M.~Kamionkowski, and K.~Griest, ``{Supersymmetric
  dark matter}'',} \textit{ Phys. Rept.} \textbf{ 267} (1996) 195,
  \href{http://dx.doi.org/10.1016/0370-1573(95)00058-5}{\doi{10.1016/0370-1573(95)00058-5}},
\href{http://www.arXiv.org/abs/hep-ph/9506380}{\texttt{arXiv:hep-ph/9506380}}.
%%CITATION = HEP-PH/9506380;%%.

\bibitem{LHC_jinst}
\hrefCMSnoop {}{E.~Lyndon and B.~Philip, ``{LHC Machine}'',} \textit{ JINST}
  \textbf{ 3} (2008) S08001,
\href{http://dx.doi.org/10.1088/1748-0221/3/08/S08001}{\doi{10.1088/1748-0221/3/08/S08001}}.
%%CITATION = JINST,3,S08001;%%.

\bibitem{Simp1}
\hrefCMSnoop {}{J.~Alwall, P.~C. Schuster, and N.~Toro, ``{Simplified models
  for a first characterization of new physics at the LHC}'',} \textit{ Phys.
  Rev. D} \textbf{ 79} (2009)
  \href{http://dx.doi.org/10.1103/PhysRevD.79.075020}{\doi{10.1103/PhysRevD.79.075020}},
\href{http://www.arXiv.org/abs/0810.3921}{\texttt{arXiv:0810.3921}}.
%%CITATION = 0810.3921;%%.

\bibitem{Simp2}
\hrefCMSnoop {}{J.~Alwall, M.-P. Le, M.~Lisanti, and J.~G. Wacker,
  ``{Model-independent jets plus missing energy searches}'',} \textit{ Phys.
  Rev. D} \textbf{ 79} (2009)
  \href{http://dx.doi.org/10.1103/PhysRevD.79.015005}{\doi{10.1103/PhysRevD.79.015005}},
\href{http://www.arXiv.org/abs/0809.3264}{\texttt{arXiv:0809.3264}}.
%%CITATION = 0809.3264;%%.

\bibitem{Simp3}
\hrefCMSnoop {}{{LHC New Physics Working Group}, D.~Alves {et~al.},
  ``{Simplified models for LHC new physics searches}'',} \textit{ J. Phys. G}
  \textbf{ 39} (2012) 105005,
  \href{http://dx.doi.org/10.1088/0954-3899/39/10/105005}{\doi{10.1088/0954-3899/39/10/105005}},
\href{http://www.arXiv.org/abs/1105.2838}{\texttt{arXiv:1105.2838}}.
%%CITATION = ARXIV.

\bibitem{SUS16008}
\hrefCMSnoop {}{{CMS Collaboration}, ``{Searches for pair production of
  third-generation squarks in $\sqrt{s}=13\TeV$ pp collisions}'',} \textit{
  Eur. Phys. J. C} \textbf{ 77} (2017) 327,
  \href{http://dx.doi.org/10.1140/epjc/s10052-017-4853-2}{\doi{10.1140/epjc/s10052-017-4853-2}},
\href{http://www.arXiv.org/abs/1612.03877}{\texttt{arXiv:1612.03877}}.
%%CITATION = ARXIV:1612.03877;%%.

\bibitem{ATLAS1}
\hrefCMSnoop {}{{ATLAS Collaboration}, ``{Search for a supersymmetric partner
  to the top quark in final states with jets and missing transverse momentum at
  $\sqrt{s}=7\TeV$ with the ATLAS detector}'',} \textit{ Phys. Rev. Lett.}
  \textbf{ 109} (2012) 211802,
  \href{http://dx.doi.org/10.1103/PhysRevLett.109.211802}{\doi{10.1103/PhysRevLett.109.211802}},
\href{http://www.arXiv.org/abs/1208.1447}{\texttt{arXiv:1208.1447}}.
%%CITATION = ARXIV:1208.1447;%%.

\bibitem{ATLAS2}
\hrefCMSnoop {}{{ATLAS Collaboration}, ``{Search for direct top squark pair
  production in final states with one isolated lepton, jets, and missing
  transverse momentum in $\sqrt{s}=7\TeV$ pp collisions using 4.7\fbinv of
  ATLAS data}'',} \textit{ Phys. Rev. Lett.} \textbf{ 109} (2012) 211803,
  \href{http://dx.doi.org/10.1103/PhysRevLett.109.211803}{\doi{10.1103/PhysRevLett.109.211803}},
\href{http://www.arXiv.org/abs/1208.2590}{\texttt{arXiv:1208.2590}}.
%%CITATION = ARXIV:1208.2590;%%.

\bibitem{ATLAS5}
\hrefCMSnoop {}{{ATLAS Collaboration}, ``{Search for a heavy top-quark partner
  in final states with two leptons with the ATLAS detector at the LHC}'',}
  \textit{ JHEP} \textbf{ 11} (2012) 094,
  \href{http://dx.doi.org/10.1007/JHEP11(2012)094}{\doi{10.1007/JHEP11(2012)094}},
\href{http://www.arXiv.org/abs/1209.4186}{\texttt{arXiv:1209.4186}}.
%%CITATION = ARXIV:1209.4186;%%.

\bibitem{ATLAS5a}
\hrefCMSnoop {}{{ATLAS Collaboration}, ``{Search for direct top-squark pair
  production in final states with two leptons in $pp$ collisions at
  $\sqrt{s}=8\TeV$ with the ATLAS detector}'',} \textit{ JHEP} \textbf{ 06}
  (2014) 124,
  \href{http://dx.doi.org/10.1007/JHEP06(2014)124}{\doi{10.1007/JHEP06(2014)124}},
\href{http://www.arXiv.org/abs/1403.4853}{\texttt{arXiv:1403.4853}}.
%%CITATION = ARXIV:1403.4853;%%.

\bibitem{ATLAS6}
\hrefCMSnoop {}{{ATLAS Collaboration}, ``{Search for direct third-generation
  squark pair production in final states with missing transverse momentum and
  two b-jets in $\sqrt{s}=8\TeV$ $pp$ collisions with the ATLAS detector}'',}
  \textit{ JHEP} \textbf{ 10} (2013) 189,
  \href{http://dx.doi.org/10.1007/JHEP10(2013)189}{\doi{10.1007/JHEP10(2013)189}},
\href{http://www.arXiv.org/abs/1308.2631}{\texttt{arXiv:1308.2631}}.
%%CITATION = ARXIV:1308.2631;%%.

\bibitem{ATLAS7}
\hrefCMSnoop {}{{ATLAS Collaboration}, ``{Measurement of Spin Correlation in
  Top-Antitop Quark Events and Search for Top Squark Pair Production in $pp$
  Collisions at $ \sqrt{s} = 8\TeV $ Using the ATLAS Detector}'',} \textit{
  Phys. Rev. Lett.} \textbf{ 114} (2015) 142001,
  \href{http://dx.doi.org/10.1103/PhysRevLett.114.142001}{\doi{10.1103/PhysRevLett.114.142001}},
\href{http://www.arXiv.org/abs/1412.4742}{\texttt{arXiv:1412.4742}}.
%%CITATION = ARXIV:1412.4742;%%.

\bibitem{ATLAS8}
\hrefCMSnoop {}{{ATLAS Collaboration}, ``{Search for pair-produced
  third-generation squarks decaying via charm quarks or in compressed
  supersymmetric scenarios in $pp$ collisions at $\sqrt{s}=8\TeV$ with the
  ATLAS detector}'',} \textit{ Phys. Rev. D} \textbf{ 90} (2014) 052008,
  \href{http://dx.doi.org/10.1103/PhysRevD.90.052008}{\doi{10.1103/PhysRevD.90.052008}},
\href{http://www.arXiv.org/abs/1407.0608}{\texttt{arXiv:1407.0608}}.
%%CITATION = ARXIV:1407.0608;%%.

\bibitem{atlas-stop1l-2015}
\hrefCMSnoop {}{{ATLAS Collaboration}, ``{ATLAS Run 1 searches for direct pair
  production of third-generation squarks at the Large Hadron Collider}'',}
  \textit{ Eur. Phys. J. C} \textbf{ 75} (2015) 510,
  \href{http://dx.doi.org/10.1140/epjc/s10052-015-3726-9}{\doi{10.1140/epjc/s10052-015-3726-9}},
  \href{http://www.arXiv.org/abs/1506.08616}{\texttt{arXiv:1506.08616}}.

\bibitem{CMS-STOP-lepton}
\hrefCMSnoop {}{{CMS Collaboration}, ``{Search for top-squark pair production
  in the single-lepton final state in pp collisions at $\sqrt{s}=8\TeV$}'',}
  \textit{ Eur. Phys. J. C} \textbf{ 73} (2013) 2677,
  \href{http://dx.doi.org/10.1140/epjc/s10052-013-2677-2}{\doi{10.1140/epjc/s10052-013-2677-2}},
\href{http://www.arXiv.org/abs/1308.1586}{\texttt{arXiv:1308.1586}}.
%%CITATION = ARXIV:1308.1586;%%.

\bibitem{CMS-alphaT}
\hrefCMSnoop {}{{CMS Collaboration}, ``{Search for supersymmetry in hadronic
  final states with missing transverse energy using the variables
  $\alpha_\mathrm{T}$ and b-quark multiplicity in pp collisions at $\sqrt{s} =
  8\TeV$}'',} \textit{ Eur. Phys. J. C} \textbf{ 73} (2013) 2568,
  \href{http://dx.doi.org/10.1140/epjc/s10052-013-2568-6}{\doi{10.1140/epjc/s10052-013-2568-6}},
\href{http://www.arXiv.org/abs/1303.2985}{\texttt{arXiv:1303.2985}}.
%%CITATION = ARXIV:1303.2985;%%.

\bibitem{RAZOR_8TeV}
\hrefCMSnoop {}{{CMS Collaboration}, ``{Search for supersymmetry using razor
  variables in events with $b$-tagged jets in $pp$ collisions at
  $\sqrt{s}=8\TeV$}'',} \textit{ Phys. Rev. D} \textbf{ 91} (2015) 052018,
  \href{http://dx.doi.org/10.1103/PhysRevD.91.052018}{\doi{10.1103/PhysRevD.91.052018}},
\href{http://www.arXiv.org/abs/1502.00300}{\texttt{arXiv:1502.00300}}.
%%CITATION = ARXIV:1502.00300;%%.

\bibitem{stop8TeV}
\hrefCMSnoop {}{{CMS Collaboration}, ``{Searches for third-generation squark
  production in fully hadronic final states in proton-proton collisions at
  $\sqrt{s}=8\TeV$}'',} \textit{ JHEP} \textbf{ 06} (2015) 116,
  \href{http://dx.doi.org/10.1007/JHEP06(2015)116}{\doi{10.1007/JHEP06(2015)116}},
\href{http://www.arXiv.org/abs/1503.08037}{\texttt{arXiv:1503.08037}}.
%%CITATION = ARXIV:1503.08037;%%.

\bibitem{stop0l_8TeV}
\hrefCMSnoop {}{{CMS Collaboration}, ``{Search for direct pair production of
  supersymmetric top quarks decaying to all-hadronic final states in pp
  collisions at $\sqrt{s}=8\TeV$}'',} \textit{ Eur. Phys. J. C} \textbf{ 76}
  (2016) 460,
  \href{http://dx.doi.org/10.1140/epjc/s10052-016-4292-5}{\doi{10.1140/epjc/s10052-016-4292-5}},
\href{http://www.arXiv.org/abs/1603.00765}{\texttt{arXiv:1603.00765}}.
%%CITATION = ARXIV:1603.00765;%%.

\bibitem{SUS15002}
\hrefCMSnoop {}{{CMS Collaboration}, ``{Search for supersymmetry in the
  multijet and missing transverse momentum final state in pp collisions at
  13\TeV}'',} \textit{ Phys. Lett. B} \textbf{ 758} (2016) 152,
  \href{http://dx.doi.org/10.1016/j.physletb.2016.05.002}{\doi{10.1016/j.physletb.2016.05.002}},
\href{http://www.arXiv.org/abs/1602.06581}{\texttt{arXiv:1602.06581}}.
%%CITATION = ARXIV:1602.06581;%%.

\bibitem{SUS15003}
\hrefCMSnoop {}{{CMS Collaboration}, ``{Search for new physics with the
  $M_{\mathrm{T2}}$ variable in all-jets final states produced in pp collisions
  at $\sqrt{s}=13\TeV$}'',} \textit{ JHEP} \textbf{ 10} (2016) 006,
  \href{http://dx.doi.org/10.1007/JHEP10(2016)006}{\doi{10.1007/JHEP10(2016)006}},
\href{http://www.arXiv.org/abs/1603.04053}{\texttt{arXiv:1603.04053}}.
%%CITATION = ARXIV:1603.04053;%%.

\bibitem{SUS15004}
\hrefCMSnoop {}{{CMS Collaboration}, ``{Inclusive search for supersymmetry
  using razor variables in $pp$ collisions at $\sqrt{s}=13\TeV$}'',} \textit{
  Phys. Rev. D} \textbf{ 95} (2017) 012003,
  \href{http://dx.doi.org/10.1103/PhysRevD.95.012003}{\doi{10.1103/PhysRevD.95.012003}},
\href{http://www.arXiv.org/abs/1609.07658}{\texttt{arXiv:1609.07658}}.
%%CITATION = ARXIV:1609.07658;%%.

\bibitem{SUS15005}
\hrefCMSnoop {}{{CMS Collaboration}, ``{A search for new phenomena in pp
  collisions at $\sqrt{s}=13\TeV$ in final states with missing transverse
  momentum and at least one jet using the ${\alpha_{\mathrm{T}}}$ variable}'',}
  \textit{ Eur. Phys. J. C} \textbf{ 77} (2017) 294,
  \href{http://dx.doi.org/10.1140/epjc/s10052-017-4787-8}{\doi{10.1140/epjc/s10052-017-4787-8}},
\href{http://www.arXiv.org/abs/1611.00338}{\texttt{arXiv:1611.00338}}.
%%CITATION = ARXIV:1611.00338;%%.

\bibitem{SUS16033}
\hrefCMSnoop {}{{CMS Collaboration}, ``{Search for supersymmetry in multijet
  events with missing transverse momentum in proton-proton collisions at
  13\TeV}'',} \textit{ Phys. Rev. D} \textbf{ 96} (2017) 032003,
  \href{http://dx.doi.org/10.1103/PhysRevD.96.032003}{\doi{10.1103/PhysRevD.96.032003}},
\href{http://www.arXiv.org/abs/1704.07781}{\texttt{arXiv:1704.07781}}.
%%CITATION = ARXIV:1704.07781;%%.

\bibitem{ATLASCom}
\hrefCMSnoop {}{{ATLAS Collaboration}, ``{Search for new phenomena in final
  states with an energetic jet and large missing transverse momentum in $pp$
  collisions at $\sqrt{s}=13\TeV$ using the ATLAS detector}'',} \textit{ Phys.
  Rev. D} \textbf{ 94} (2016) 032005,
  \href{http://dx.doi.org/10.1103/PhysRevD.94.032005}{\doi{10.1103/PhysRevD.94.032005}},
\href{http://www.arXiv.org/abs/1604.07773}{\texttt{arXiv:1604.07773}}.
%%CITATION = ARXIV:1604.07773;%%.

\bibitem{ATLAS2b}
\hrefCMSnoop {}{{ATLAS Collaboration}, ``{Search for bottom squark pair
  production in proton-proton collisions at $\sqrt{s}=13$ TeV with the ATLAS
  detector}'',} Technical Report~10, 2016.
\newblock
  \href{http://dx.doi.org/10.1140/epjc/s10052-016-4382-4}{\doi{10.1140/epjc/s10052-016-4382-4}},
  \href{http://www.arXiv.org/abs/1606.08772}{\texttt{arXiv:1606.08772}}.

\bibitem{ATLASJet}
\hrefCMSnoop {}{{ATLAS Collaboration}, ``{Search for squarks and gluinos in
  final states with jets and missing transverse momentum at $\sqrt{s}=13\TeV$
  with the ATLAS detector}'',} \textit{ Eur. Phys. J. C} \textbf{ 76} (2016)
  392,
  \href{http://dx.doi.org/10.1140/epjc/s10052-016-4184-8}{\doi{10.1140/epjc/s10052-016-4184-8}},
\href{http://www.arXiv.org/abs/1605.03814}{\texttt{arXiv:1605.03814}}.
%%CITATION = ARXIV:1605.03814;%%.

\bibitem{CDF1}
\hrefCMSnoop {}{{CDF} Collaboration, ``{Search for 3- and 4-body decays of the
  scalar top quark in $\text{p} \bar{\text{p}}$ collisions at $\sqrt{s} =
  1.8\TeV$}'',} \textit{ Phys. Lett. B} \textbf{ 581} (2004) 147,
  \href{http://dx.doi.org/10.1016/j.physletb.2003.12.001}{\doi{10.1016/j.physletb.2003.12.001}}.

\bibitem{CDF2}
\hrefCMSnoop {}{{CDF} Collaboration, ``{Search for the supersymmetric partner
  of the top quark in $\text{p} \bar{\text{p}}$ collisions at $\sqrt{s} =
  1.96\TeV$}'',} \textit{ Phys. Rev. D} \textbf{ 82} (2010) 092001,
  \href{http://dx.doi.org/10.1103/PhysRevD.82.092001}{\doi{10.1103/PhysRevD.82.092001}},
\href{http://www.arXiv.org/abs/1009.0266}{\texttt{arXiv:1009.0266}}.
%%CITATION = ARXIV:1009.0266;%%.

\bibitem{D01}
\hrefCMSnoop {}{{D0} Collaboration, ``{Search for the lightest scalar top quark
  in events with two leptons in $\text{p} \bar{\text{p}}$ collisions at
  $\sqrt{s} = 1.96\TeV$}'',} \textit{ Phys. Lett. B} \textbf{ 659} (2008) 500,
  \href{http://dx.doi.org/10.1016/j.physletb.2007.11.086}{\doi{10.1016/j.physletb.2007.11.086}},
\href{http://www.arXiv.org/abs/0707.2864}{\texttt{arXiv:0707.2864}}.
%%CITATION = ARXIV:0707.2864;%%.

\bibitem{D02}
\hrefCMSnoop {}{{D0} Collaboration, ``{Search for pair production of the scalar
  top quark in muon+tau final states}'',} \textit{ Phys. Lett. B} \textbf{ 710}
  (2012) 578,
  \href{http://dx.doi.org/10.1016/j.physletb.2012.03.028}{\doi{10.1016/j.physletb.2012.03.028}},
\href{http://www.arXiv.org/abs/1202.1978}{\texttt{arXiv:1202.1978}}.
%%CITATION = ARXIV:1202.1978;%%.

\bibitem{Trigger}
\hrefCMSnoop {}{{CMS Collaboration}, ``{The CMS trigger system}'',} \textit{
  JINST} \textbf{ 12} (2017) P01020,
  \href{http://dx.doi.org/10.1088/1748-0221/12/01/P01020}{\doi{10.1088/1748-0221/12/01/P01020}},
\href{http://www.arXiv.org/abs/1609.02366}{\texttt{arXiv:1609.02366}}.
%%CITATION = ARXIV:1609.02366;%%.

\bibitem{CMS_jinst}
\hrefCMSnoop {}{{CMS Collaboration}, ``The {CMS} experiment at the {CERN}
  {LHC}'',} \textit{ JINST} \textbf{ 3} (2008) S08004,
\href{http://dx.doi.org/10.1088/1748-0221/3/08/S08004}{\doi{10.1088/1748-0221/3/08/S08004}}.
%%CITATION = JINST,3,S08004;%%.

\bibitem{PF}
\hrefCMSnoop {}{{CMS Collaboration}, ``{Particle-flow reconstruction and global
  event description with the CMS detector}'',} \textit{ JINST} \textbf{ 12}
  (2017) P10003,
  \href{http://dx.doi.org/10.1088/1748-0221/12/10/P10003}{\doi{10.1088/1748-0221/12/10/P10003}},
\href{http://www.arXiv.org/abs/1706.04965}{\texttt{arXiv:1706.04965}}.
%%CITATION = ARXIV:1706.04965;%%.

\bibitem{Cacciari:2008gp}
\hrefCMSnoop {}{M.~Cacciari, G.~P. Salam, and G.~Soyez, ``{The anti-$k_t$ jet
  clustering algorithm}'',} \textit{ JHEP} \textbf{ 04} (2008) 063,
  \href{http://dx.doi.org/10.1088/1126-6708/2008/04/063}{\doi{10.1088/1126-6708/2008/04/063}},
  \href{http://www.arXiv.org/abs/0802.1189}{\texttt{arXiv:0802.1189}}.

\bibitem{Cacciari:2011ma}
\hrefCMSnoop {}{M.~Cacciari, G.~P. Salam, and G.~Soyez, ``{FastJet user
  manual}'',} \textit{ Eur. Phys. J. C} \textbf{ 72} (2012) 1896,
  \href{http://dx.doi.org/10.1140/epjc/s10052-012-1896-2}{\doi{10.1140/epjc/s10052-012-1896-2}},
\href{http://www.arXiv.org/abs/1111.6097}{\texttt{arXiv:1111.6097}}.
%%CITATION = ARXIV:1111.6097;%%.

\bibitem{Jet2}
\hrefCMSnoop {}{M.~Cacciari and G.~P. Salam, ``Pileup subtraction using jet
  areas'',} \textit{ Phys. Lett. B} \textbf{ 659} (2007) 119,
  \href{http://dx.doi.org/10.1016/j.physletb.2007.09.077}{\doi{10.1016/j.physletb.2007.09.077}},
  \href{http://www.arXiv.org/abs/0707.1378}{\texttt{arXiv:0707.1378}}.

\bibitem{JER}
\hrefCMSnoop {}{{CMS Collaboration}, ``{Jet energy scale and resolution in the
  CMS experiment in pp collisions at 8\TeV}'',} \textit{ JINST} \textbf{ 12}
  (2017) P02014,
  \href{http://dx.doi.org/10.1088/1748-0221/12/02/P02014}{\doi{10.1088/1748-0221/12/02/P02014}},
\href{http://www.arXiv.org/abs/1607.03663}{\texttt{arXiv:1607.03663}}.
%%CITATION = ARXIV:1607.03663;%%.

\bibitem{Met1}
\hrefCMSnoop {}{{CMS Collaboration}, ``{Performance of the CMS missing
  transverse momentum reconstruction in pp data at $\sqrt{s}=8\TeV$}'',}
  \textit{ JINST} \textbf{ 10} (2015) P02006,
  \href{http://dx.doi.org/10.1088/1748-0221/10/02/P02006}{\doi{10.1088/1748-0221/10/02/P02006}},
\href{http://www.arXiv.org/abs/1411.0511}{\texttt{arXiv:1411.0511}}.
%%CITATION = ARXIV:1411.0511;%%.

\bibitem{Muon}
\hrefCMSnoop {}{{CMS Collaboration}, ``{Performance of CMS muon reconstruction
  in pp collision events at $\sqrt{s} = 7\TeV$}'',} \textit{ JINST} \textbf{ 7}
  (2012) P10002,
  \href{http://dx.doi.org/10.1088/1748-0221/7/10/P10002}{\doi{10.1088/1748-0221/7/10/P10002}},
\href{http://www.arXiv.org/abs/1206.4071}{\texttt{arXiv:1206.4071}}.
%%CITATION = ARXIV:1206.4071;%%.

\bibitem{Khachatryan:2015hwa}
\hrefCMSnoop {}{{CMS Collaboration}, ``{Performance of electron reconstruction
  and selection with the CMS detector in proton-proton collisions at
  $\sqrt{s}=8\TeV$}'',} \textit{ JINST} \textbf{ 10} (2015) P06005,
  \href{http://dx.doi.org/10.1088/1748-0221/10/06/P06005}{\doi{10.1088/1748-0221/10/06/P06005}},
\href{http://www.arXiv.org/abs/1502.02701}{\texttt{arXiv:1502.02701}}.
%%CITATION = ARXIV:1502.02701;%%.

\bibitem{CMS-PAS-JME-14-001}
\href {https://cds.cern.ch/record/1751454}{{CMS Collaboration}, ``{Study of
  pileup removal algorithms for jet}'',} CMS Physics Analysis Summary
  CMS-PAS-JME-14-001, 2014.

\bibitem{Btags}
\hrefCMSnoop {}{{CMS Collaboration}, ``{Identification of b-quark jets with the
  CMS experiment}'',} \textit{ JINST} \textbf{ 8} (2013) P04013,
  \href{http://dx.doi.org/10.1088/1748-0221/8/04/P04013}{\doi{10.1088/1748-0221/8/04/P04013}},
\href{http://www.arXiv.org/abs/1211.4462}{\texttt{arXiv:1211.4462}}.
%%CITATION = ARXIV:1211.4462;%%.

\bibitem{BTV1}
\href {https://cds.cern.ch/record/2138504}{{CMS Collaboration},
  ``{Identification of b quark jets at the CMS experiment in the LHC Run 2}'',}
  CMS Physics Analysis Summary CMS-PAS-BTV-15-001, 2016.

\bibitem{ctag}
\href {http://cds.cern.ch/record/2205149}{{CMS Collaboration},
  ``{Identification of c-quark jets at the CMS experiment}'',} CMS Physics
  Analysis Summary CMS-PAS-BTV-16-001, 2016.

\bibitem{Khachatryan:2011wq}
\hrefCMSnoop {}{{CMS Collaboration}, ``{Measurement of $\mathrm{ B\overline{B}
  }$ angular correlations based on secondary vertex reconstruction at
  $\sqrt{s}=7\TeV$}'',} \textit{ JHEP} \textbf{ 03} (2011) 136,
  \href{http://dx.doi.org/10.1007/JHEP03(2011)136}{\doi{10.1007/JHEP03(2011)136}},
\href{http://www.arXiv.org/abs/1102.3194}{\texttt{arXiv:1102.3194}}.
%%CITATION = ARXIV:1102.3194;%%.

\bibitem{CMS-SUS-16-049}
\hrefCMSnoop {}{{CMS Collaboration}, ``{Search for direct production of
  supersymmetric partners of the top quark in the all-jets final state in
  proton-proton collisions at $\sqrt{s} = 13\TeV$}'',} \textit{ JHEP} \textbf{
  10} (2017) 005,
  \href{http://dx.doi.org/10.1007/JHEP10(2017)005}{\doi{10.1007/JHEP10(2017)005}},
  \href{http://www.arXiv.org/abs/1707.03316}{\texttt{arXiv:1707.03316}}.

\bibitem{Alwall:2014hca}
J.~Alwall\hrefCMSnoop {}{ {et~al.}, ``{The automated computation of tree-level
  and next-to-leading order differential cross sections, and their matching to
  parton shower simulations}'',} \textit{ JHEP} \textbf{ 07} (2014) 079,
  \href{http://dx.doi.org/10.1007/JHEP07(2014)079}{\doi{10.1007/JHEP07(2014)079}},
\href{http://www.arXiv.org/abs/1405.0301}{\texttt{arXiv:1405.0301}}.
%%CITATION = ARXIV:1405.0301;%%.

\bibitem{Ball:2014uwa}
\hrefCMSnoop {}{{NNPDF} Collaboration, ``{Parton distributions for the LHC Run
  II}'',} \textit{ JHEP} \textbf{ 04} (2015) 040,
  \href{http://dx.doi.org/10.1007/JHEP04(2015)040}{\doi{10.1007/JHEP04(2015)040}},
\href{http://www.arXiv.org/abs/1410.8849}{\texttt{arXiv:1410.8849}}.
%%CITATION = ARXIV:1410.8849;%%.

\bibitem{Nason:2004rx}
\hrefCMSnoop {}{P.~Nason, ``{A new method for combining NLO QCD with shower
  Monte Carlo algorithms}'',} \textit{ JHEP} \textbf{ 11} (2004) 040,
  \href{http://dx.doi.org/10.1088/1126-6708/2004/11/040}{\doi{10.1088/1126-6708/2004/11/040}},
\href{http://www.arXiv.org/abs/hep-ph/0409146}{\texttt{arXiv:hep-ph/0409146}}.
%%CITATION = HEP-PH/0409146;%%.

\bibitem{Frixione:2007vw}
\hrefCMSnoop {}{S.~Frixione, P.~Nason, and C.~Oleari, ``{Matching NLO QCD
  computations with parton shower simulations: the POWHEG method}'',} \textit{
  JHEP} \textbf{ 11} (2007) 070,
  \href{http://dx.doi.org/10.1088/1126-6708/2007/11/070}{\doi{10.1088/1126-6708/2007/11/070}},
\href{http://www.arXiv.org/abs/0709.2092}{\texttt{arXiv:0709.2092}}.
%%CITATION = ARXIV:0709.2092;%%.

\bibitem{Alioli:2010xd}
\hrefCMSnoop {}{S.~Alioli, P.~Nason, C.~Oleari, and E.~Re, ``{A general
  framework for implementing NLO calculations in shower Monte Carlo programs:
  the POWHEG BOX}'',} \textit{ JHEP} \textbf{ 06} (2010) 043,
  \href{http://dx.doi.org/10.1007/JHEP06(2010)043}{\doi{10.1007/JHEP06(2010)043}},
\href{http://www.arXiv.org/abs/1002.2581}{\texttt{arXiv:1002.2581}}.
%%CITATION = ARXIV:1002.2581;%%.

\bibitem{Re:2010bp}
\hrefCMSnoop {}{E.~Re, ``{Single-top $Wt$-channel production matched with
  parton showers using the POWHEG method}'',} \textit{ Eur. Phys. J. C}
  \textbf{ 71} (2011) 1547,
  \href{http://dx.doi.org/10.1140/epjc/s10052-011-1547-z}{\doi{10.1140/epjc/s10052-011-1547-z}},
\href{http://www.arXiv.org/abs/1009.2450}{\texttt{arXiv:1009.2450}}.
%%CITATION = ARXIV:1009.2450;%%.

\bibitem{Sjostrand:2014zea}
T.~Sj{\"o}strand\hrefCMSnoop {}{ {et~al.}, ``An introduction to {PYTHIA}
  8.2'',} \textit{ Comput. Phys. Commun.} \textbf{ 191} (2015) 159,
  \href{http://dx.doi.org/10.1016/j.cpc.2015.01.024}{\doi{10.1016/j.cpc.2015.01.024}},
\href{http://www.arXiv.org/abs/1410.3012}{\texttt{arXiv:1410.3012}}.
%%CITATION = ARXIV:1410.3012;%%.

\bibitem{geant4}
\hrefCMSnoop {}{{{GEANT4}} Collaboration, ``{GEANT4---a simulation toolkit}'',}
  \textit{ Nucl. Instr. Meth. A} \textbf{ 506} (2003) 250,
\href{http://dx.doi.org/10.1016/S0168-9002(03)01368-8}{\doi{10.1016/S0168-9002(03)01368-8}}.
%%CITATION = NUIMA,A506,250;%%.

\bibitem{fastsim}
S.~Abdullin\hrefCMSnoop {}{ {et~al.}, ``The fast simulation of the {CMS}
  detector at {LHC}'',} \textit{ J. Phys. Conf. Ser.} \textbf{ 331} (2011)
  032049,
\href{http://dx.doi.org/10.1088/1742-6596/331/3/032049}{\doi{10.1088/1742-6596/331/3/032049}}.
%%CITATION = 00462,331,032049;%%.

\bibitem{Borschensky:2014cia}
C.~Borschensky\hrefCMSnoop {}{ {et~al.}, ``{Squark and gluino production cross
  sections in $pp$ collisions at $\sqrt{s} = $ 13, 14, 33 and 100\TeV}'',}
  \textit{ Eur. Phys. J. C} \textbf{ 74} (2014) 3174,
  \href{http://dx.doi.org/10.1140/epjc/s10052-014-3174-y}{\doi{10.1140/epjc/s10052-014-3174-y}},
\href{http://www.arXiv.org/abs/1407.5066}{\texttt{arXiv:1407.5066}}.
%%CITATION = ARXIV:1407.5066;%%.

\bibitem{MCT1}
\hrefCMSnoop {}{D.~R. Tovey, ``{On measuring the masses of pair-produced
  semi-invisibly decaying particles at hadron colliders}'',} \textit{ JHEP}
  \textbf{ 04} (2008) 034,
  \href{http://dx.doi.org/10.1088/1126-6708/2008/04/034}{\doi{10.1088/1126-6708/2008/04/034}},
\href{http://www.arXiv.org/abs/0802.2879}{\texttt{arXiv:0802.2879}}.
%%CITATION = ARXIV:0802.2879;%%.

\bibitem{MCT}
\hrefCMSnoop {}{G.~Polesello and D.~R. Tovey, ``{Supersymmetric particle mass
  measurement with the boost-corrected contransverse mass}'',} \textit{ JHEP}
  \textbf{ 03} (2010) 030,
  \href{http://dx.doi.org/10.1007/JHEP03(2010)030}{\doi{10.1007/JHEP03(2010)030}},
\href{http://www.arXiv.org/abs/0910.0174}{\texttt{arXiv:0910.0174}}.
%%CITATION = ARXIV:0910.0174;%%.

\bibitem{PDG}
\hrefCMSnoop {}{{Particle Data Group}, C.~Patrignani {et~al.}, ``{Review of
  Particle Physics}'',} \textit{ Chin. Phys. C} \textbf{ 40} (2016) 100001,
\href{http://dx.doi.org/10.1088/1674-1137/40/10/100001}{\doi{10.1088/1674-1137/40/10/100001}}.
%%CITATION = CHPHD,C38,090001;%%.

\bibitem{CMS-PAS-LUM-17-001}
\href {https://cds.cern.ch/record/2257069}{{CMS Collaboration}, ``{CMS
  luminosity measurements for the 2016 data taking period}'',} CMS Physics
  Analysis Summary CMS-PAS-LUM-17-001, 2017.

\bibitem{CLs1}
\hrefCMSnoop {}{T.~Junk, ``Confidence level computation for combining searches
  with small statistics'',} \textit{ Nucl. Instr. Meth. A} \textbf{ 434} (1999)
  435,
  \href{http://dx.doi.org/10.1016/S0168-9002(99)00498-2}{\doi{10.1016/S0168-9002(99)00498-2}},
  \href{http://www.arXiv.org/abs/hep-ex/9902006}{\texttt{arXiv:hep-ex/9902006}}.

\bibitem{CLs}
\hrefCMSnoop {}{A.~L. Read, ``{Presentation of search results: the CLs
  technique}'',} \textit{ J. Phys. G} \textbf{ 28} (2002) 2693,
  \href{http://dx.doi.org/10.1088/0954-3899/28/10/313}{\doi{10.1088/0954-3899/28/10/313}}.

\bibitem{combine}
\href {http://cdsweb.cern.ch/record/1379837}{{ATLAS and CMS Collaborations},
  ``{Procedure for the LHC Higgs boson search combination in Summer 2011}'',}
  Technical Report ATL-PHYS-PUB-2011-11, CMS-NOTE-2011-005, 2011.

\bibitem{asy}
\hrefCMSnoop {}{G.~Cowan, K.~Cranmer, E.~Gross, and O.~Vitells, ``{Asymptotic
  formulae for likelihood-based tests of new physics}'',} \textit{ Eur. Phys.
  J. C} \textbf{ 71} (2011) 1554,
  \href{http://dx.doi.org/10.1140/epjc/s10052-011-1554-0}{\doi{10.1140/epjc/s10052-011-1554-0}},
  \href{http://www.arXiv.org/abs/1007.1727}{\texttt{arXiv:1007.1727}}.
[Erratum \DOI{10.1140/epjc/s10052-013-2501-z}].
%%CITATION = ARXIV:1007.1727;%%.

\bibitem{REInt}
\href {http://cds.cern.ch/record/2242860}{{CMS Collaboration}, ``{Simplified
  likelihood for the re-interpretation of public CMS results}'',} Technical
  Report CERN-CMS-NOTE-2017-001, 2017.

\end{thebibliography}\endgroup

\clearpage
\newpage
\numberwithin{table}{section}
\numberwithin{figure}{section}
\appendix

\section{Correlation matrices for background estimates}
\label{sec:app_result}

\begin{figure*}[!ht]
\centering
\includegraphics[width=0.70\textwidth]{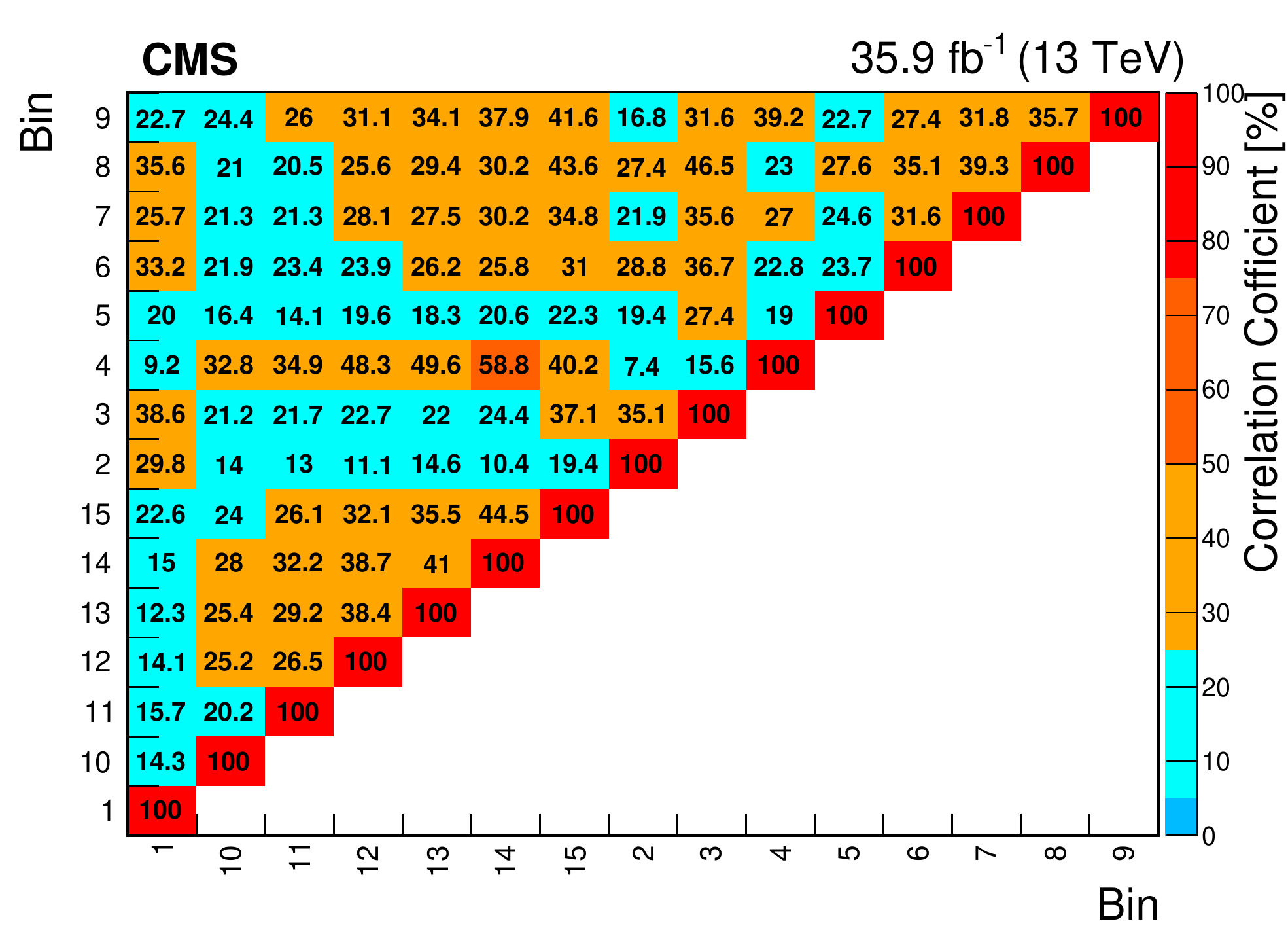} \\
\caption{The correlation matrix for the estimated backgrounds in the noncompressed search region. The bin numbers are defined in Table~\ref{tab:noncom_all}.
\label{fig:Cov}}
\end{figure*}

\begin{figure*}[!ht]
\centering
\includegraphics[width=0.70\textwidth]{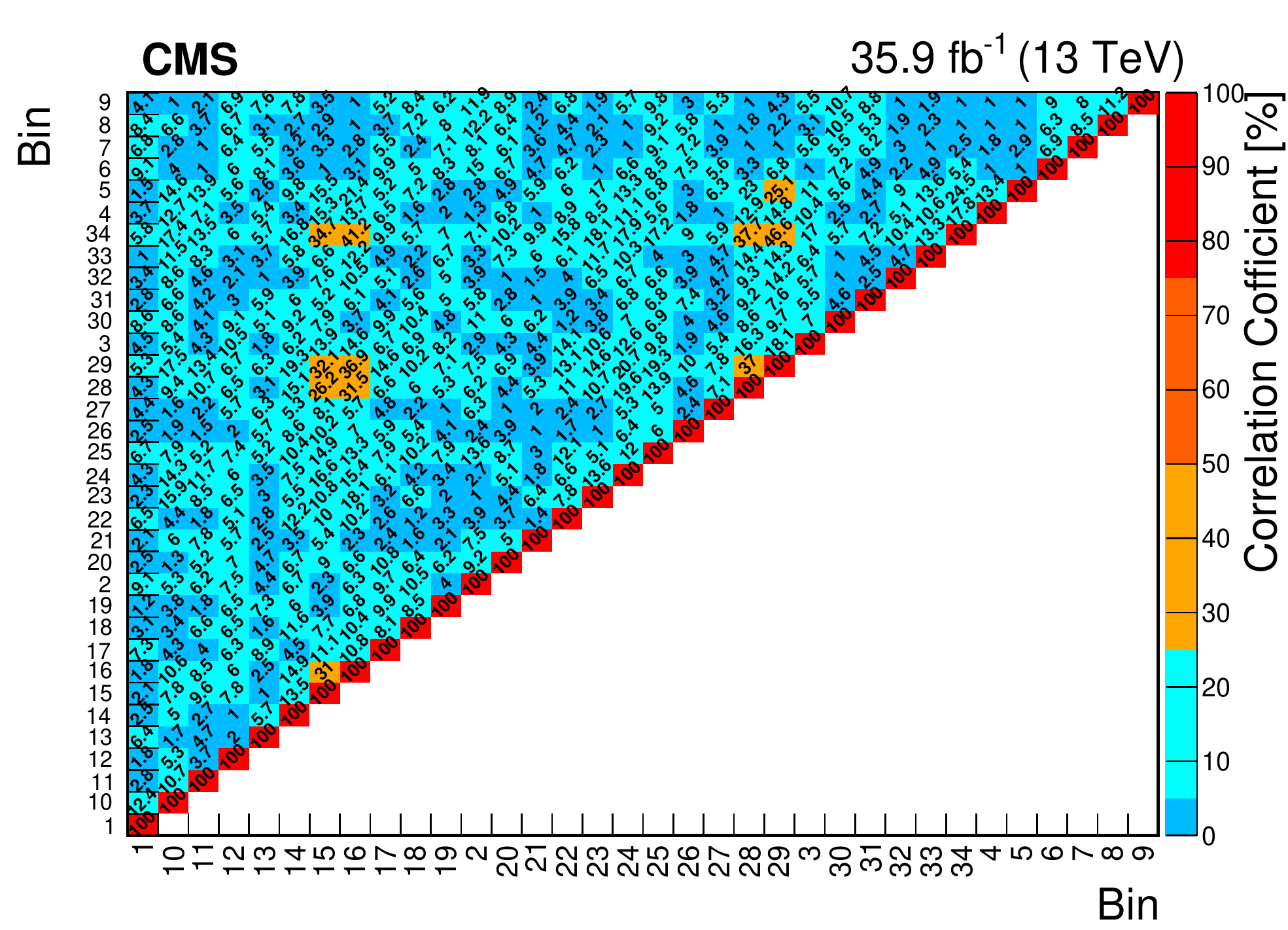}\\
\caption{The correlation matrix for the estimated backgrounds in the compressed search region. The bin numbers are defined in Table~\ref{tab:com_bin}. \label{fig:Cov1}}
\end{figure*}

To facilitate reinterpretation of the results in a broader range of beyond the standard model scenarios \cite{REInt}, the correlation matrices for the background estimates in the noncompressed and compressed search regions are provided in Figs.~\ref{fig:Cov} and~\ref{fig:Cov1}, respectively. The bin number in the compressed region is the same as in Table~\ref{tab:noncom_all} of our paper and in the noncompressed region shown below in Table~\ref{tab:com_bin}.

\begin{table*}[!htp]
\topcaption{ The bin number and definition for the compressed search region as shown in Fig.~\ref{fig:Cov} above.
\label{tab:com_bin}}
\centering
\begin{tabular}{c c c c}
\multicolumn{4}{c}{Compressed region}    \\ \hline
\Nb, \Nc, \NI&    \met [\GeVns{}] & \HT (\PQb- or \PQc-tagged jets) [\GeVns{}] & Bin   \\ \hline
\multirow{5}{*}{\Nb = 1}& 250--300 &  $<$100 & 1 \\
& 300--500 &  $<$100 & 2 \\
& 500--750 &  $<$100 & 3 \\
& \x750--1000 &  $<$100 & 4 \\
& $>$1000 & $<$100 & 5\\ \hline
\multirow{6}{*}{\Nb = 2}&  \multirow{2}{*}{250--300} & $<$100 &6  \\
&  & 100--200 & 7 \\
& \multirow{2}{*}{300--500} & $<$100 & 8 \\
& & 100--200 &  9 \\
& \multirow{2}{*}{$>$500} & $<$100 & 10 \\
& & 100--200 &  11 \\ \hline
\multirow{5}{*}{\Nc = 1}& 250--300 &  $<$100 & 12 \\
& 300--500 &  $<$100 & 13 \\
& 500--750 &  $<$100 & 14 \\
& \x750--1000 &  $<$100 & 15 \\
& $>$1000 & $<$100 & 16 \\ \hline
\multirow{8}{*}{\Nc = 2}&  \multirow{2}{*}{250--300} & $<$100 & 17 \\
&  & 100--200 &  18 \\
& \multirow{2}{*}{300--500} & $<$100 & 19 \\
& & 100--200 &   20 \\
& \multirow{2}{*}{500--750} & $<$100 & 21 \\
& & 100--200 & 22 \\
& \multirow{2}{*}{$>$750} & $<$100 & 23 \\
& & 100--200 &  24 \\ \hline
\multirow{5}{*}{$\Nb+ \Nc = 0$, $\NI > 0$}& 250--300 & --- & 25 \\
& 300--500 & --- & 26 \\
& 500--750 & --- & 27  \\
& \x750--1000 & --- & 28 \\
& $>$1000 & --- & 29 \\ \hline
\multirow{5}{*}{\Nb+ \Nc+ \NI$ =$ 0}& 300--500 & --- & 30 \\
& 500--750 & --- & 31 \\
& \x750--1000 & --- & 32 \\
& 1000--1250 & --- & 33  \\
& $>$1250 & --- & 34 \\
\end{tabular}
\end{table*}

\clearpage

\cleardoublepage \section{The CMS Collaboration \label{app:collab}}\begin{sloppypar}\hyphenpenalty=5000\widowpenalty=500\clubpenalty=5000\textbf{Yerevan Physics Institute,  Yerevan,  Armenia}\\*[0pt]
A.M.~Sirunyan, A.~Tumasyan
\vskip\cmsinstskip
\textbf{Institut f\"{u}r Hochenergiephysik,  Wien,  Austria}\\*[0pt]
W.~Adam, F.~Ambrogi, E.~Asilar, T.~Bergauer, J.~Brandstetter, E.~Brondolin, M.~Dragicevic, J.~Er\"{o}, M.~Flechl, M.~Friedl, R.~Fr\"{u}hwirth\cmsAuthorMark{1}, V.M.~Ghete, J.~Grossmann, J.~Hrubec, M.~Jeitler\cmsAuthorMark{1}, A.~K\"{o}nig, N.~Krammer, I.~Kr\"{a}tschmer, D.~Liko, T.~Madlener, I.~Mikulec, E.~Pree, D.~Rabady, N.~Rad, H.~Rohringer, J.~Schieck\cmsAuthorMark{1}, R.~Sch\"{o}fbeck, M.~Spanring, D.~Spitzbart, J.~Strauss, W.~Waltenberger, J.~Wittmann, C.-E.~Wulz\cmsAuthorMark{1}, M.~Zarucki
% ...
\vskip\cmsinstskip
% ...
1:~~Also at Vienna University of Technology, Vienna, Austria\\
% ...

\end{sloppypar}
\end{document}